\documentclass{aa}  
\usepackage{comment}
\usepackage{graphicx}
\usepackage{txfonts}
\usepackage{multirow}
\usepackage{subcaption}
\usepackage{lscape}          
\usepackage{adjustbox}
\usepackage{amssymb}
\usepackage{placeins}   
\usepackage{xcolor}
\usepackage{colortbl}
\usepackage{linenoaa}
\usepackage[colorlinks=true,allcolors=blue]{hyperref}

\makeatletter
\renewcommand*\aa@pageof{, page \thepage{} of \pageref*{LastPage}}
\makeatother

\newcommand{\lya}{Ly$\alpha$}

\usepackage{textgreek}
\usepackage{xargs}
\usepackage{xcolor}
\usepackage{xspace}

\let\oldtextsigma\textsigma
\renewcommand{\textsigma}{\oldtextsigma\xspace}
\let\oldAA\AA
\renewcommand{\AA}{\text{\oldAA}\xspace}
\def\w80{\ensuremath{w_{80}}\xspace}

\newcommandx{\fluxdcgs}[1][1=-20]{$\times 10^{[#1]}$~erg~s$^{-1}$~cm$^{-2}$~\AA$^{-1}$\xspace}

\newcommand{\Halpha}{\text{H\textalpha}\xspace}
\newcommand{\Hbeta}{\text{H\textbeta}\xspace}
\newcommand{\Hgamma}{\text{H\textgamma}\xspace}
\newcommand{\Hdelta}{\text{H\textdelta}\xspace}
\newcommand{\Heta}{\text{H\texteta}\xspace}
\newcommand{\Hzeta}{\text{H\textzeta}\xspace}
\newcommandx{\permittedEL}[6][1=O,2=III,3=,4=,5=,6=]{\text{{#1}\,{\sc{#2}}{#3}{#4}{#5}{#6}}\xspace}
\newcommandx{\semiforbiddenEL}[6][1=O,2=III,3=,4=,5=,6=]{\text{{#1}\,{\sc{#2}}]{#3}{#4}{#5}{#6}}\xspace}
\newcommandx{\forbiddenEL}[6][1=O,2=III,3=,4=,5=,6=]{\text{[{#1}\,{\sc{#2}}]{#3}{#4}{#5}{#6}}\xspace}

\newcommandx{\HI}{\permittedEL[H][i]}
\newcommandx{\HII}{\permittedEL[H][ii]}
\newcommandx{\HeI}{\permittedEL[He][i]}
\newcommandx{\HeIL}[1][1=3889]{\permittedEL[He][i][\,\textlambda][#1]}
\newcommandx{\HeIIL}[1][1=4686]{\permittedEL[He][ii][\,\textlambda][#1]}

\newcommandx{\HeII}{\permittedEL[He][ii]}
\newcommand{\OIII}{\forbiddenEL[O][iii]}
\newcommandx{\OIIIuv}[1][1=1666]{\semiforbiddenEL[O][iii][\textlambda][#1]}
\newcommandx{\OIIIL}[1][1=5007]{\forbiddenEL[O][iii][\textlambda][#1]}
\newcommand{\OIIIall}{\forbiddenEL[O][iii][\textlambda][\textlambda][4959,][5007]}
\newcommandx{\NIL}[1]{\forbiddenEL[N][i][\textlambda][5200]}

\newcommandx{\OIL}[1][1=8446]{\permittedEL[O][i][\textlambda][#1]}
\newcommandx{\OILs}[1][1=8446]{\text{\textlambda {#1}}\xspace}

\newcommand{\OII}{\forbiddenEL[O][ii]}
\newcommandx{\OIIL}[1][1=3727]{\forbiddenEL[O][ii][\textlambda][#1]}
\newcommand{\OIIall}{\forbiddenEL[O][ii][\textlambda][\textlambda][][3726,3729]}
\newcommand{\NeIIIall}{\forbiddenEL[Ne][iii][\textlambda][\textlambda][][3869,3967]}

\newcommand{\NeIII}{\forbiddenEL[Ne][iii][\textlambda][3869]}
\newcommand{\NeIIIe}{\forbiddenEL[Ne][iii]}

\newcommand{\NII}{\forbiddenEL[N][ii][\textlambda][6584]}
\newcommand{\NIIall}{\forbiddenEL[N][ii][\textlambda][\textlambda][][6548,6584]}

\newcommandx{\NIIL}[1][1=6583]{\forbiddenEL[N][ii][\textlambda][#1]}

\newcommandx{\CIV}{\permittedEL[C][iv]}

\newcommandx{\CaT}{\permittedEL[Ca][ii][\textlambda][][8498,8542,8662]}
\newcommandx{\CaL}[1][1=8498]{\permittedEL[Ca][ii][\textlambda][#1]}
\newcommandx{\Ca}{\permittedEL[Ca][ii]}

\newcommandx{\Fef}{\forbiddenEL[Fe][ii]}

\newcommandx{\Fe}{\permittedEL[Fe][ii]}
\newcommandx{\FeL}{\permittedEL[Fe][ii][\textlambda]}

\newcommandx{\Feopt}{\permittedEL[Fe][ii][\textlambda][\textlambda][5190,][5320]}

\newcommandx{\CIII}{\semiforbiddenEL[C][iii]}

\newcommandx{\MgII}{\permittedEL[Mg][ii]}

\newcommand{\SII}{\forbiddenEL[S][ii]}

\newcommand{\ArIII}{\forbiddenEL[Ar][iii]}

\begin{document}

	\title{The Collective Voice of Ly$\alpha$ Emitters: Insights from JWST Stacked Spectroscopy}
	
	\titlerunning{Properties of LAEs}
	\authorrunning{R. Tripodi et al.}
	
\author{Roberta~Tripodi
    \inst{1,2}\thanks{\email{roberta.tripodi@inaf.it}}
    \and Lorenzo~Napolitano
    \inst{1}
    \and Laura~Pentericci
    \inst{1}
    \and Borja~P\'{e}rez-D\'{\i}az
    \inst{1}
    \and Aniket~Bhagwat 
    \inst{3}
    \and Francesco~D'Eugenio
    \inst{4,5}
    \and Flor~Arévalo-González
    \inst{1}
    \and Pablo~G.~Pérez-González
    \inst{6}
    \and Antonio~Arroyo-Polonio
    \inst{1}
    \and Antonello~Calabrò
    \inst{1}
    \and Benedetta~Ciardi
    \inst{3}
    \and Mark~Dickinson
    \inst{7}
    \and Henry~C.~Ferguson
    \inst{8}
    \and Giovanni~Gandolfi
    \inst{1}
    \and Michaela~Hirschmann
    \inst{9}
    \and Weida~Hu
    \inst{10,11}
    \and A.~M.~Koekemoer
    \inst{8}
    \and Mario~Llerena
    \inst{1}
    \and Ray~A.~Lucas
    \inst{8}
    \and M.~S.~Oey
    \inst{12}
    \and Casey~Papovich
    \inst{10,11}
    \and {L.~Y.~Aaron}~{Yung}
    \inst{8}
    \and Xin~Wang
    \inst{13,14,15}
}

 \institute{INAF - Osservatorio Astronomico di Roma, via Frascati 33, I-00078, Monte Porzio Catone, Italy 
         \and
         IFPU - Institute for Fundamental Physics of the Universe, via Beirut 2, I-34151 Trieste, Italy 
    \and
    \text{Max Planck Institut für Astrophysik, Karl Schwarzschild Straße 1, D-85741 Garching, Germany} 
    \and
    Kavli Institute for Cosmology, University of Cambridge, Madingley Road,
Cambridge CB3 0HA, UK 
    \and
    Cavendish Laboratory – Astrophysics Group, University of Cambridge, 19
JJ Thomson Avenue, Cambridge CB3 0HE, UK 
    \and 
    Centro de Astrobiologia (CAB), CSIC-INTA, Ctra. de Ajalvir km 4, Torrejon de Ardoz, E-28850, Madrid, Spain 
    \and NSF's National Optical-Infrared Astronomy Research Laboratory, 950 N. Cherry Ave., Tucson, AZ 85719, USA 
    \and
    Space Telescope Science Institute, 3700 San Martin Drive, Baltimore, MD 21218, USA 
    \and Institute of Physics, Laboratory of Galaxy Evolution, Ecole Polytechnique F{\'e}d{\'e}rale de Lausanne (EPFL), Observatoire de Sauverny, 1290 Versoix, Switzerland 
    \and
    Department of Physics and Astronomy, Texas A\&M University, College Station, TX 77843-4242, USA 
    \and
    George P. and Cynthia Woods Mitchell Institute for Fundamental Physics and Astronomy, Texas A\&M University, College Station, TX 77843-4242, USA 
    \and
    Astronomy Department, University of Michigan, Ann Arbor, MI 48109-1107, USA 
    \and
    School of Astronomy and Space Science, University of Chinese Academy of Sciences (UCAS), Beijing 100049, China 
    \and
    National Astronomical Observatories, Chinese Academy of Sciences, Beijing 100101, China 
    \and
    Institute for Frontiers in Astronomy and Astrophysics, Beijing Normal University, Beijing 102206, China 
    }

	\abstract{How Ly$\alpha$ photons escape from galaxies during the first billion years remains a central question for understanding early galaxy evolution and cosmic reionization. While Ly$\alpha$ emitters (LAEs) are known to be metal-poor and highly ionized, many high-z studies rely on spatially integrated measurements, leaving the internal connection between Ly$\alpha$ emission, chemical enrichment, and feedback largely unexplored. We present a spatially resolved stacking analysis of 287 LAEs at $z>4$ observed with JWST/NIRSpec prism spectroscopy. By constructing a two-dimensional stack from public surveys (CAPERS, CEERS, JADES, and RUBIES), we probe the average internal structure of typical LAEs ($z_{\rm median}=5.5$, $M_{\rm UV, median}=-18.7$, EW(\lya)$_{\rm median}$ = 61~\AA) on sub-kiloparsec scales. We find a clear radial decoupling between resonant and non-resonant emission: while EW(\Hbeta) and other optical lines decline with radius, EW(Ly$\alpha$) increases toward the outskirts, and the Ly$\alpha$ escape fraction rises from $\sim16\%$ in the center to $\gtrsim24\%$ at larger radii. This behavior suggests that resonant scattering redistributes Ly$\alpha$ photons into outer regions, where escape becomes more efficient. Optical diagnostics and $T_e$ measurements reveal low metallicities ($12+\log(\rm O/H)\simeq7.7\pm0.2$), high ionization parameters, negligible dust attenuation, and systematically elevated N/O ratios ($\log({\rm N/O})\sim-0.4$). The latter place typical LAEs among the growing population of nitrogen-enhanced high-redshift galaxies, pointing to rapid and possibly feedback-driven chemical enrichment. The inferred ionizing photon production efficiency, $\log(\xi_{\rm ion}/{\rm Hz\,erg^{-1}})\simeq25.1-25.2$, together with the high Ly$\alpha$ escape fractions, suggests that these systems are efficient, though not extreme, contributors to the ionizing photon budget. Comparison with SPICE radiation-hydrodynamic simulations shows that bursty supernova feedback models naturally reproduce the observed radial trends in Ly$\alpha$ escape, UV slope, and emission-line equivalent widths, linking the spatial redistribution of Ly$\alpha$ to stochastic star formation and feedback-driven gas flows. Our results demonstrate that Ly$\alpha$ emission, chemical enrichment, and feedback are tightly connected in typical $z>4$ LAEs, and highlight the power of spatially resolved stacking with JWST to uncover the internal physical drivers of early galaxy evolution.}

	
	\maketitle
	
	\section{Introduction}
	\label{sec:intro}
	
	\nolinenumbers
	
	\begin{figure}
		\centering
		\includegraphics[width=0.9\linewidth]{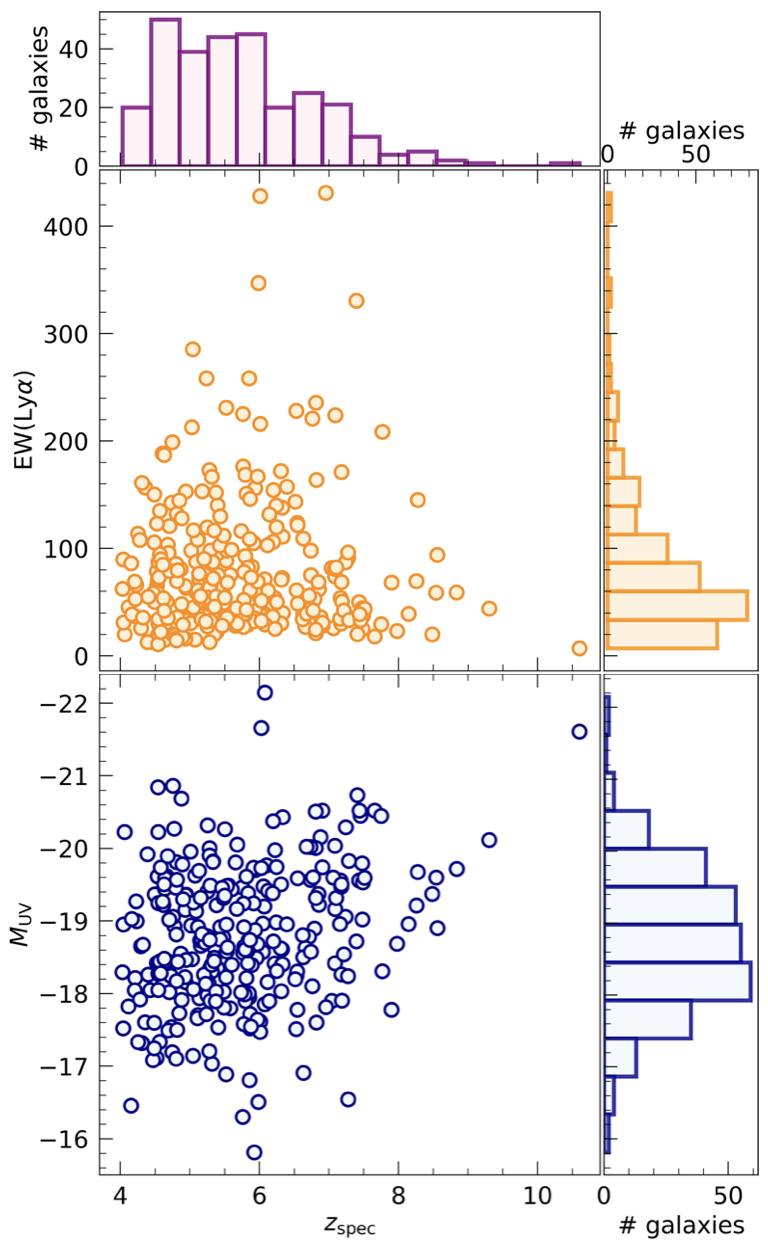}
		\caption{From top to bottom: EW(\lya), $M_{\rm UV}$ vs redshift in our sample of 287 LAEs. Histograms reporting the distributions of rest-frame EW(\lya), $M_{\rm UV}$, and redshift are also shown at the side of the main panels.}
		\label{fig:sample}
	\end{figure}
	
	\begin{figure*}
		\centering
		\includegraphics[width=1\linewidth]{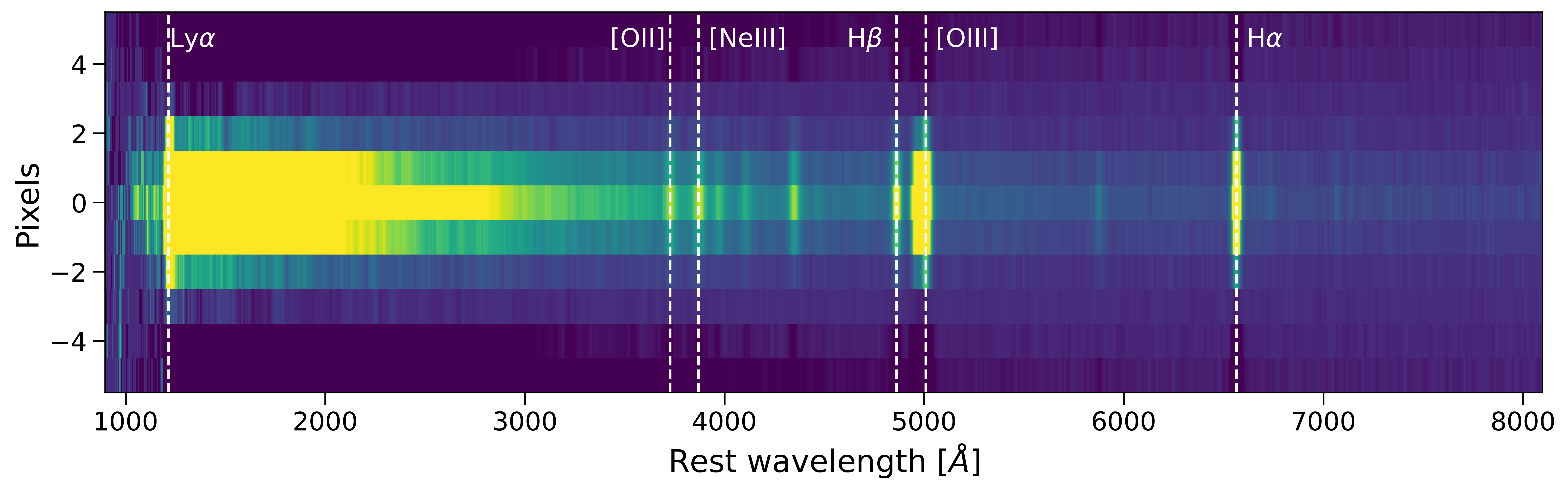}
		\caption{2D stacked spectrum of 287 LAEs using the NIRSpec/MSA prism spectra of LAEs at $z>4$. One pixel corresponds to 0.1 arcsec on the y-axis. Some of the brightest detected emission lines are marked with a vertical dashed white line.}
		\label{fig:stack}
	\end{figure*}
	
	\begin{figure*}
		\centering
		\includegraphics[width=0.9\linewidth]{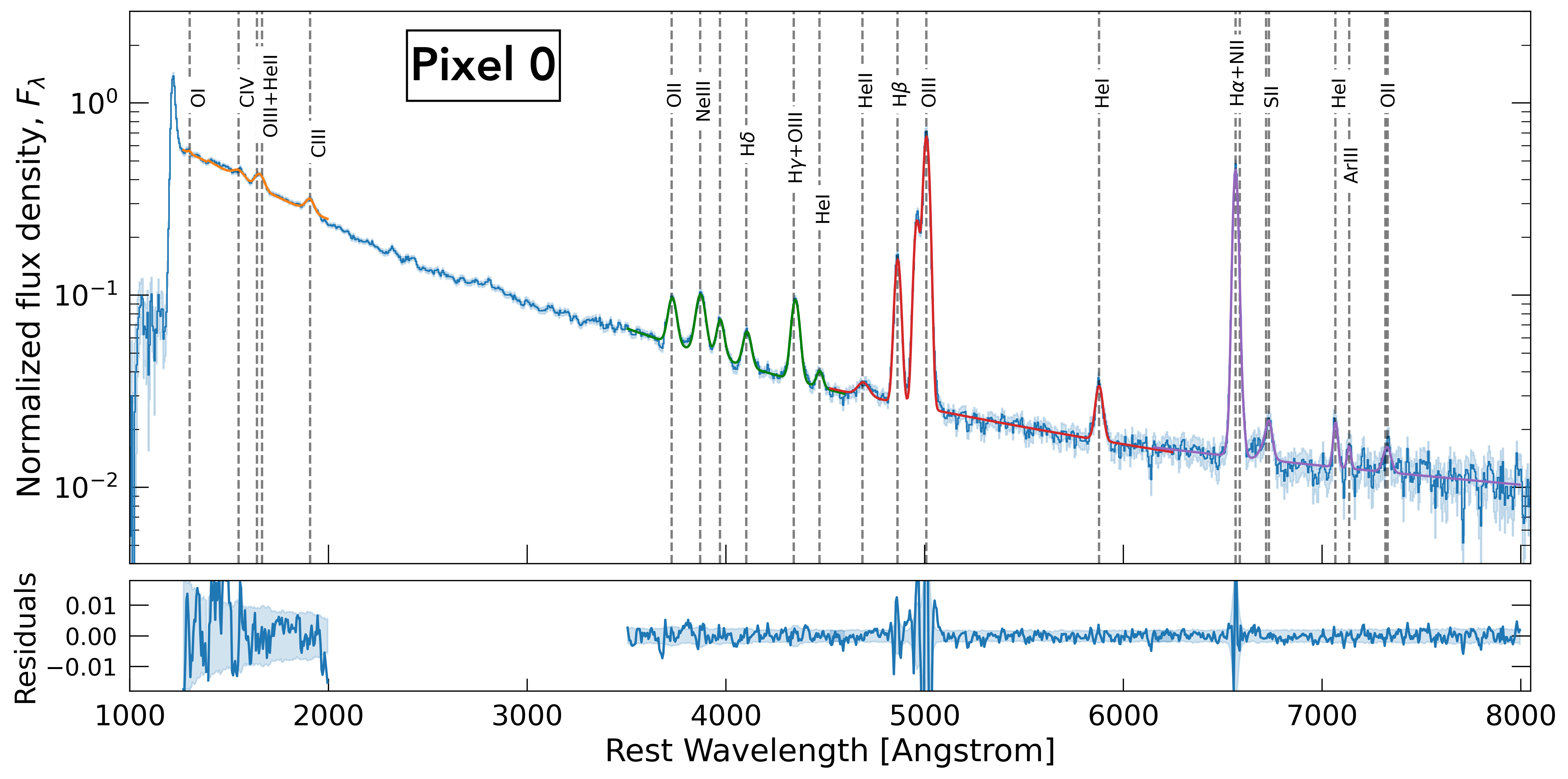}
		\caption{Spectrum extracted from pixel 0 of the LAEs stack. (Top) Best-fit models to the observed data (blue solid line) are shown as solid colored lines. Different colors mark different wavelength groups for which the fit is performed independently, as described in Sect. \ref{sec:method}. (Bottom) Residuals of the best-fitting models shown in the top panel. 1$\sigma$ noise level is reported as shaded blue region.}
		\label{fig:fit-spec-pix0}
	\end{figure*}
	
	\lya ~emitters (LAEs) represent a fundamental population of galaxies in the early Universe. Characterized by strong \lya ~emission and typically faint UV continua, LAEs are thought to trace low-mass, metal-poor systems with intense star formation \citep[e.g.,][]{Kornei2010,Verhamme2018, Ono2010, ouchi2020}. Because \lya ~photons are resonantly scattered by neutral hydrogen, their escape depends sensitively on the geometry, kinematics of the gas, as well as on the geometry of the dust \citep{verhamme2006}. As a result, LAEs provide a unique laboratory for studying the interplay between star formation, interstellar medium (ISM) properties, and radiative transfer at high redshift \citep[e.g.,][]{kusakabe2022, Jung2024}.
	
	Over the past decade, large photometric and spectroscopic surveys have established LAEs as a key feature found in the galaxy demographic at $z>4$, spanning the epoch of reionization and its immediate aftermath.  Their low metallicities, blue UV slopes, and high ionization parameters suggest that LAEs may represent an early evolutionary phase of star-forming events, potentially linked to efficient production of  ionizing photons \citep{wilkins2011,cullen2023,Napolitano2023}. The presence of strong \lya\ emission is in general an indication of non negligible escape of Lyman continuum photons \citep[e.g.,][]{marchi2018, pahl2024, dijkstra2016, izotov2018, gazagnes2020, naidu2022} and, for this reason, LAEs are also thought to be key contributors to cosmic reionization \citep[e.g.,][]{saxena2024, matthee2022}. However, most observational constraints on LAEs have relied on spatially integrated measurements, limiting our ability to connect \lya ~emission to the internal physical conditions of galaxies.

	Understanding the spatial origin of \lya ~emission is particularly critical. Observations have shown that \lya ~halos can be significantly more extended than the UV continuum, suggesting that resonant scattering redistributes \lya ~photons over large spatial scales and that \lya ~photons may also be scattered by the circum-galactic medium \citep{leclercq2017, kusakabe2022, hayes2014, steidel2011}. At higher redshifts, direct spatially resolved studies are far more challenging due to the intrinsic faintness and compactness of LAEs. As a consequence, it has been unclear how \lya ~emission relates to the distribution of stars, metals, and ionized gas within early galaxies, and whether spatial gradients in metallicity, ionization, or dust play a dominant role in regulating \lya ~escape.
	
	The advent of the James Webb Space Telescope \citep[JWST,][]{gardner2006, gargner2023} has transformed this landscape. In particular, NIRSpec prism observations provide continuous spectral coverage from the rest-frame UV to the optical at $z>4$, enabling simultaneous access to \lya, UV metal lines, Balmer recombination lines, and key optical diagnostics such as \OIIall, \OIIIL, and \NIIall. Although the prism mode operates at low spectral resolution ($R\simeq30$–300), its unprecedented sensitivity allows the detection of faint emission lines in galaxies that were previously accessible only through narrow-band \lya ~surveys \citep[e.g.,][]{Jones2024B, Kageura2025, Tang2024, yamada2012,inoue2020,umeda2025}.
	
	Early JWST observations have already revealed the remarkable diversity of physical conditions in high-redshift galaxies, including extreme ionization states, low metallicities, and signatures of chemical enrichment that challenge simple evolutionary models. At the same time, JWST has enabled the first systematic detection of rest-frame optical lines in large samples of LAEs, opening a new window onto their ionized gas properties. However, even with JWST, spatially resolved spectroscopy of individual LAEs remains feasible only through deep observations, with the most notable case being GN-z11 $z=10.6$ which shows  a 0.4$''$ \lya-UV peak offset and a $\sim$0.78$''$ extended \lya\ profile \citep{Maiolino2024b, Scholtz2024, bunker2023}.
	
	Stacking techniques therefore provide a powerful complementary approach to extract key spectral features and infer physical properties across objects of similar nature. By co-adding spectra from large samples of galaxies, stacking enables measurements of average properties that would otherwise be inaccessible \citep{roberts-borsani2024, glazer2025}. When applied to two-dimensional spectra, 2D spectral stacking can preserve spatial information, allowing radial trends to be studied statistically even when individual objects are too faint to be well detected \citep{Tripodi2024c}. 
	
	In this work, we exploit the growing archive of JWST/NIRSpec prism observations to perform a spatially resolved stacking analysis of LAEs at $z>4$. We compile a sample of 287 LAEs drawn from multiple public surveys across the COSMOS, EGS, UDS, and GOODS-N/S fields, making this one of the largest spectroscopic samples of high-redshift LAEs assembled to date. By stacking, we achieve high signal-to-noise ratios while retaining radial information on scales of $\sim0.1''$, corresponding to sub-kiloparsec physical distances at the median redshift of the sample.
	
	This stacking strategy allows us to investigate, in a uniform and self-consistent framework, how \lya ~emission relates to UV continuum slopes, Balmer recombination lines, metal-line diagnostics, and chemical abundances as a function of radius. In particular, we use a combination of empirical diagnostics, photoionization models, and direct electron-temperature measurements to constrain metallicity, ionization parameter, and nitrogen enrichment. We further estimate \lya ~escape fractions and the ionizing photon production efficiency, $\xi_{\rm ion}$, and explore how these quantities vary spatially within the stacked galaxies.
	
	By combining the statistical power of stacking with the broad wavelength coverage of JWST, this work provides a population-level view of the internal structure of LAEs during the first billion years of cosmic history. Our results offer new insights into the physical mechanisms governing \lya ~escape, chemical enrichment, and ionizing photon production in early galaxies, and demonstrate the effectiveness of stacked JWST spectroscopy as a tool for studying systems beyond the reach of individual observations.
	
	Throughout the paper we adopt AB magnitudes \citep{oke1983} and the $\Lambda$CDM cosmology from \citet{planck2018}: $H_0=67.4 ~\rm km ~s^{-1} ~Mpc^{-1}$, $\Omega_m=0.315$, and $\Omega_{\Lambda}=0.685$. We assume Case~B recombination with electron temperature $T_\mathrm{e} = 10{,}000$~K and electron density $n_\mathrm{e} = 10^3~\mathrm{cm}^{-3}$ \citep{storey+hummer1995}, unless otherwise specified.

	\section{A sample of Ly$\alpha$ emitters at $z>4$}
	\label{sec:sample}
	
	The aim of this study is to investigate the connection between \lya ~emission and rest-frame UV and optical properties of \lya ~emitters (LAEs) at $z > 4$ using spatially resolved spectroscopy. For this purpose, JWST/NIRSpec Multi Shutter Array (MSA) prism observations offer the optimal combination of broad spectral coverage ($\sim$0.6–5.3 $\mu$m) and high sensitivity, despite their relatively low spectral resolution ($R \simeq 30$–300). We compiled all publicly available NIRSpec/prism spectroscopy from publicly available surveys  (CEERS, JADES and RUBIES\footnote{CEERS: PID 1345, PI: Steven Finkelstein \citep{backhaus2024}. JADES: PID 1180, PI: Daniel J. Eisenstein \citep{deugenio2025}. RUBIES: PID 4233, PIs: A. de Graaff \& G. Brammer \citep{deGraaff2025}.}) in the DAWN JWST Archive
	\footnote{\url{https://dawn-cph.github.io/dja/index.html}}\citep[DJA,][]{Heintz2024dja, deGraaff2025}, covering the COSMOS \citep{Scoville2007}, EGS \citep{davis2007}, UDS \citep{Lawrence2007}, and GOODS-N/S fields \citep{grogin2011, koekemoer2011}. 
	
	In particular, for the GOODS-N/S fields we considered the catalog of 82 LAEs identified by \cite{Jones2025}. Similarly, for the EGS and UDS fields we adopted the LAEs samples reported by \cite{Napolitano2024} and \cite{Napolitano2026}, comprising 42 and 74 LAEs, respectively. We then complemented these samples with new LAE identifications from the CANDELS-Area Prism Epoch of Reionization Survey (CAPERS; GO-6368, PI Mark Dickinson). 
	CAPERS observed a total of 21 NIRSpec pointings, uniformly distributed across the COSMOS, UDS, and EGS fields, using the prism configuration. For CAPERS targets, we used the internal CAPERS data reduction (see Appendix \ref{app:capers} for a comparison with DJA reduction) which follows the general procedures outlined in \cite{ArrabalHaro2023}. Details of the survey strategy, target selection, data reduction, and redshift identification are provided in the collaboration papers \citep[e.g.,][]{Napolitano2026, Taylor2025}. 
	
	Background subtraction was performed using local background estimates from adjacent shutters. As a result, some degree of self-subtraction may affect the 2D spectra of the most spatially extended sources at distances beyond $\pm2.5$ off-centered pixels. In our analysis, however, we restrict the measurements to pixels within $\pm2$, and therefore our results are not expected to be significantly biased by potential self-subtraction effects.
	
	Whenever available, we also considered spectroscopic redshifts and the absolute UV magnitude (M$_{\mathrm{UV}}$) provided by literature. For unpublished sources, we derived the redshift and M$_{\mathrm{UV}}$ from the line centroids of rest-frame optical emission lines and the median observed flux averaged over 1450--1550~\AA, respectively \citep[see][for details]{Napolitano2026}. 
	
	We note that for the identification of additional LAEs in CAPERS, we applied the same methodology adopted in the aforementioned studies \citep[][]{Napolitano2024, Jones2025, Napolitano2026}. Briefly, due to the low spectral resolution of the 1D NIRSpec prism data, \lya\ emission is modeled within a $\sim$ 4--5 pixels window centered on the observed emission peak at the expected \lya\ wavelength. As a first step, the line flux is computed via direct integration over this window. The continuum is estimated through a linear fit to the red side of \lya, spanning from 1900~\AA\ to three pixels redward of the emission peak. To further refine the measurement, we adopt a forward-modeling approach. We construct a library of Gaussian emission profiles with full widths at half maximum (FWHM) uniformly sampled in the range 100--1500 km s$^{-1}$. We note that star-forming galaxies at z $\sim$ 5--7 were found to have \lya\ FWHM $\sim$ 300 km s$^{-1}$ \citep[e.g.,][]{Pentericci2018, Tang2024}. The assumed range of FWHM has been adopted to sample properties of star-forming LAEs with a broad prior. For each profile, the amplitude is drawn from a uniform distribution constrained to produce an integrated flux within a factor of 0.2--5 of that obtained from direct integration. Each Gaussian profile is combined with a step-function continuum: the red side is given by the linear fit, while the blue side is fixed to the median flux blueward of the line, accounting for intergalactic medium absorption. The full model is convolved with a Gaussian kernel ($\sigma_R (\lambda_{\mathrm{obs}})$), matching the instrumental resolution. The fit is performed using a Markov Chain Monte Carlo (MCMC) analysis implemented with the \texttt{emcee} package \citep{foreman2013}, using 10 walkers and 20,000 steps. The best-fit \lya\ flux and rest-frame equivalent width (EW$_0$, hereafter just EW) are derived from the posterior medians, with uncertainties given by the 68th percentile intervals. Overall, following this procedure, we identified 83 and 87 additional LAEs in the CAPERS-EGS and CAPERS-COSMOS fields, respectively. 
	
	By selecting sources with signal-to-noise ratio (S/N) on the EW of \lya\ $>3$, we considered a total of 368 LAEs at z $>$ 4 across the COSMOS, EGS, UDS, and GOODS-N/S fields. All spectra were visually inspected to ensure data quality, and the following additional selection criteria were applied to guarantee the robustness of the analysis. First, to minimize AGN contamination, we excluded known broad-line AGNs reported in the literature \citep[e.g.,][]{maiolino24, juodzbalis2026}, as well as objects identified through visual inspection\footnote{We evaluate by visual inspection whether two components are required to reproduce the \Halpha profile, while the \OIII is modeled by a single narrow component.}, removing 24 sources ($\sim 6\%$ of the initial sample). Second, we discarded spectra affected by strong artifacts, including multiple spurious spikes in the spectral region of interest or severe truncation due to the physical gap between the NIRSpec detectors. Third, we excluded LAEs whose spectra were contaminated by nearby secondary sources. The final sample comprises 287 LAEs at $z > 4$.
	
	Fig. \ref{fig:sample} presents the distribution of the final sample in rest-frame equivalent width of \lya\ (EW(\lya)), $M_{\mathrm{UV}}$, and redshift. The sample has a median EW$_{\rm median}$(\lya)$=61$~\AA and a median $M_{\mathrm{UV, median}} = -18.7$, with 16th and 84th percentiles at $M_{\mathrm{UV}} = -17.9$ and $-19.7$, respectively. The median redshift of the sample is $z_{\rm median}=5.5$, with 16th and 84th percentiles at $z=4.6$ and $=6.8$, respectively. This corresponds to a median angular scale of $6.113 ~\rm kpc/"$, which we adopt as the reference distance to angular conversion scale, hereafter.

	\begin{figure}
		\centering
		\includegraphics[width=0.9\linewidth]{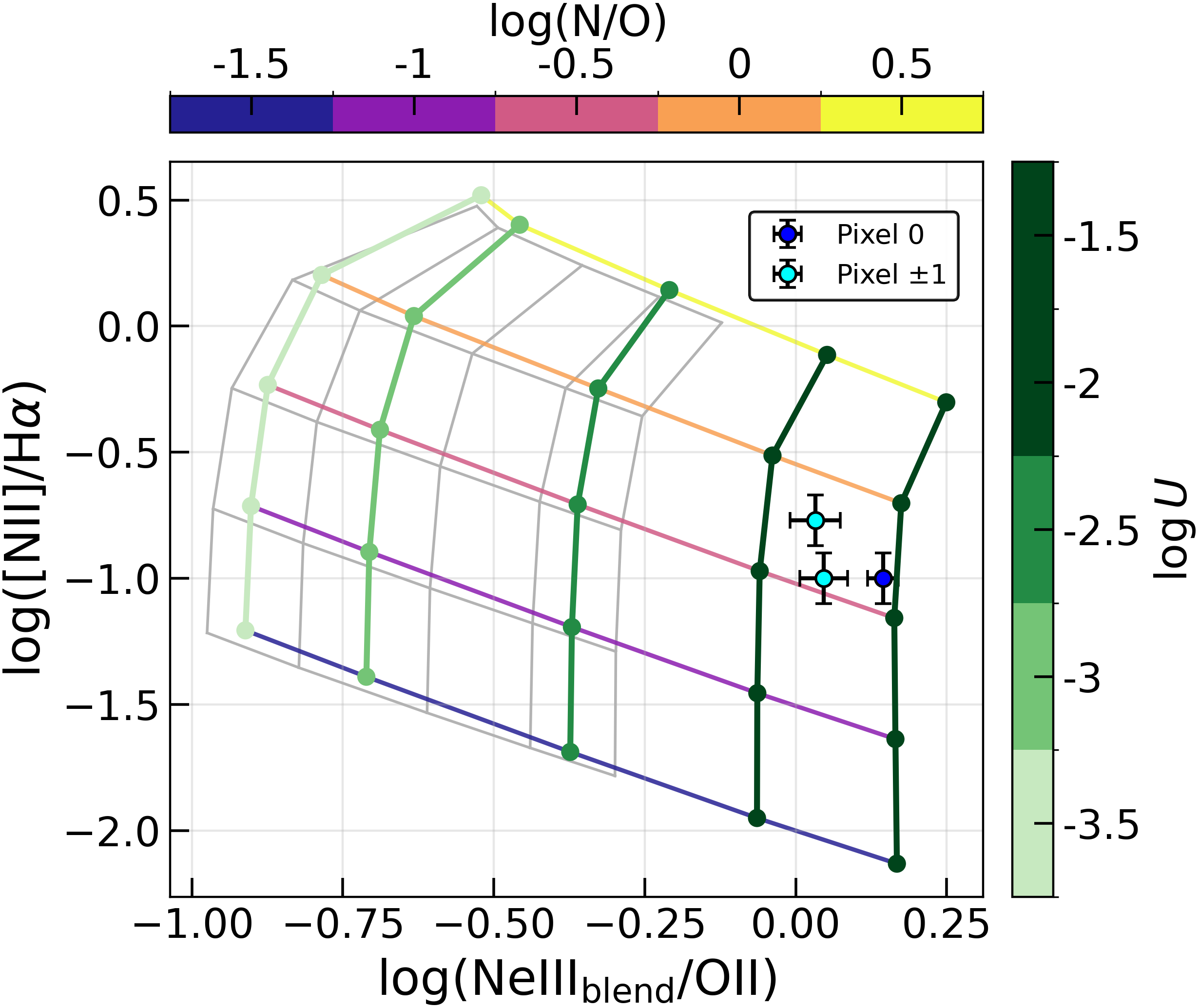}
		\caption{\NII/\Halpha vs \NeIIIe$_{\rm blend}$/\OII. Our observed ratios in pixels $0$, $\pm1$ are compared with models from the NUVOLOSO project (P\'{e}rez-D\'{\i}az et al. in prep.) at $\log(n)[\rm cm^{-3}]=2.0$ (gray scale) and $\log(n)[\rm cm^{-3}]=3.0$ (colored scale). The grid is color-coded by discrete values of ionization parameter, $\log(U)$, and nitrogen abundance, $\log(\rm N/O)$, as reported in the respective color bars.}
		\label{fig:N2Ha-ratio}
	\end{figure}

	\section{Methods}
	\label{sec:method}
	
	\begin{table*}[]
		\centering
		\caption{EW$_0$ of the fitted emission lines.}
		\begin{tabular}{c|ccccc}
			Line & Pix -2 & Pix -1 & Pix 0 & Pix 1 & Pix 2  \\
			\hline
			\lya & $63\pm 5$ & $33\pm 1$ & $30\pm 1$ & $35\pm 1$ & $74\pm 6$ \\
			\ion{C}{II}$^\dagger$ & -- & -- & -- & $1.3\pm 0.7$ & -- \\
			\ion{Si}{IV}$^\dagger$ & -- & -- & -- & $2\pm 1$ & -- \\
			\ion{C}{IV} & -- & -- & $7 \pm 1 $ & $7\pm 2$ & -- \\
			\OIIIuv+\ion{He}{ii} & -- & $8\pm 1$ &  $10\pm 2$ & $21\pm 3$ & -- \\
			\ion{C}{III}] & $18\pm 5$ & $17\pm 3$ &  $9\pm 1$ & $20\pm 2$ &  $16\pm 4$\\
			\HeIL[3487] & -- & -- & -- & $9\pm 2$ & -- \\
			\OIIall & $45\pm8$ & $49\pm 2$ & $38 \pm 1$ & $49\pm2$ & $80\pm 8$\\
			\NeIII+\Hzeta+\Heta+\HeIL[3889] & $57\pm 7$ & $60\pm 3$ & $60\pm 1 $ & $57\pm3$ & $68\pm 8$\\
			\Hdelta & $27\pm 6$ & $24\pm 2$ &  $26 \pm 1$ & $19\pm2$ & $26\pm 5$\\
			\Hgamma + \OIIIL[4363] & $76\pm 12$ & $65\pm 7$  &  $83 \pm 2 $ & $71\pm6$ & $95\pm 15$\\
			\HeIL[4471] & -- & $6\pm 2$ &  $9\pm 1$ & $4\pm 2$ & --\\
			\ion{He}{II} & -- & $4\pm 2$ &  $12 \pm 2$ & $8\pm 2$ & --\\
			\Hbeta & $144\pm10$ & $160\pm 5$ & $175 \pm 3 $ & $155 \pm4$ & $165\pm 10$\\
			\OIIIL & $716\pm 23$ & $847\pm14$ & $938 \pm 12 $ & $846\pm 13$ & $709\pm 26$\\
			\HeIL[5876] & $71\pm 15$ & $47\pm 5$ &  $41 \pm 3$ & $31\pm3$ & --\\
			\Halpha & $<722$ & $775_{-74}^{+34}$ & $837 \pm 30 $ & $769\pm 46$ & $<892$\\[0.1cm]
			[\ion{N}{II}] & -- & $76_{-21}^{+55}$& $83 \pm 20 $ & $129\pm 33$ & --\\[0.1cm]
			\SII & -- & $22\pm6$ &  $30\pm 5$ & $27\pm 6$ & --\\
			\HeIL[7065] & -- & $15\pm 5$ &  $20 \pm 3$ & $17\pm 4$ & --\\
			\ArIII$^\dagger$ & -- & -- &  $7 \pm 3 $ & $9\pm 3$ & --$^{\dagger\dagger}$\\ 
			\OII\textlambda\textlambda7319,7330 & -- & -- & $16 \pm 5$ & -- & --\\
			\hline
			\hline
		\end{tabular}
		\tablefoot{Columns: line name, EW$_0$ in different pixels from -2 to 2 in units of \AA. The symbol $^\dagger$ marks tentative detections. $^{\dagger\dagger}$: we find a tentative detection of \ArIII in pix 2, but the continuum level is too noisy to derive a reliable estimate of its EW.}
		\label{tab:ew-lines}
	\end{table*}
	
	\subsection{Stacking procedure}
	\label{sec:stack}
	
	To maximize the S/N across the spectrum, we performed a stacking analysis of the parent sample of 287 LAEs (see Sect. \ref{sec:sample}). Specifically, we constructed a composite two-dimensional (2D) spectrum of LAEs at $z>4$ by stacking their JWST/NIRSpec prism observations. Prior to stacking, each individual spectrum was de-redshifted, interpolated onto a common rest-frame wavelength grid, and normalized by the rest-frame UV continuum flux at $\lambda_{\rm rest}=1500~\AA$ derived from $M_{\rm UV}$ (see Sect.~\ref{sec:sample}). All spectra were spatially aligned with respect to the galaxy center, defined as follows by finding the peak of the radial flux distribution in continuum and emission line. Specifically, we first computed the sum of the flux over the whole spectral wavelength range for each spatial pixel; then, we fitted this spatial flux profile with a Gaussian, and we adopted the value of the peak pixel as the galaxy center. This is a commonly established method for deriving the center of a source in 2D spectra \citep[see e.g. appendix A in][]{castellano2025}, and the results have been verified by visual inspection. This alignment is essential for obtaining reliable radial profiles of both line and continuum emission.
	
	The composite flux in each wavelength and spatial bin was computed as the median of the flux distribution across the stacked spectra. We chose to adopt the median stacking over the average stacking for two main reasons: (1) to allow a fair comparison with results from other studies in the literature \citep[e.g.,][]{roberts-borsani2024, isobe2025}; (2) the median approach is known to mitigate the contribution of outliers, producing spectra that better represent the average properties of individual galaxies. Associated uncertainties were estimated as the semi-difference between the 16th and 84th percentiles of the flux distribution in each bin (i.e. (84th-16th)/2), further divided by the square root of the number of stacked galaxies. We verified that this approach yields the most conservative uncertainty estimates when compared to standard error propagation or to the dispersion measured from 1000 stacked spectra generated by randomly perturbing individual spectra within their errors \citep[see also][]{roberts-borsani2024,isobe2025}. Therefore, percentile-based uncertainties simultaneously capture the statistical error of the individual spectra and the intrinsic variation in physical properties across the galaxy population, which we intentionally preserve when modeling the stacked spectrum and inferring physical parameters.
	
	The final 2D stacked spectrum is shown in Fig.~\ref{fig:stack}. The five central spatial pixels ($0,~\pm1,~\pm2$) exhibit significant continuum and emission-line signal, while in pixels $\pm3$ only a tentative Ly$\alpha$ detection is observed (see Sect.~\ref{sec:lya} for discussion). We note that the angular distance between adjacent spatial pixels is 0.1$"$ \citep[][]{jakobsen+2022, Boker2023}.
	
	\subsection{Spectral modeling}
	\label{sec:spec-model}
	
	From the stacked 2D spectrum, we extracted one-dimensional (1D) spectra for each spatial pixel in the range $[-3,3]$ (see Fig.~\ref{fig:stack}) and fitted them independently. Emission lines and continuum were modeled using a Bayesian framework based on MCMC sampling, implemented with the \texttt{emcee} package \citep{foreman2013}. 
	Given the large wavelength coverage of each spectrum, the fitting was performed over separate wavelength intervals, simultaneously modeling emission lines and continuum within each interval (see, e.g., the colored fits in Fig.~\ref{fig:fit-spec-pix0}). The continuum was described by a power-law model ($F_\lambda \propto \lambda^{\alpha}$), with free slope and normalization. All emission lines, except \lya\ and blended lines, were modeled with a single Gaussian profile, with free peak flux, central wavelength, and velocity dispersion, yielding posterior distributions for all parameters. 
	To model blends of unrelated species (i.e., \OIIIuv+\ion{He}{II}, \Hgamma+\OIIIL[4363]), we adopted two Gaussians, as this provides a better representation of the observed profile than a single Gaussian; however, due to the strong degeneracy of the fit, we do not attempt to decompose the individual components and instead consider only the total blended flux.
	For emission-line doublets with fixed wavelength separation and intrinsic flux ratios, the number of free parameters was reduced by fixing the relative peak fluxes and wavelength offsets to their atomic values and enforcing a common line width (e.g., \NeIIIall, \NIIall, \OIIIall; see \citealt{Tripodi2024c} for details).
	
	Line fluxes, widths, and rest-frame equivalent widths were derived from the median (50th percentile) of the posterior distributions, with uncertainties estimated as the semi-difference between the 16th and 84th percentiles. The EW of all detected and tentative emission lines are reported in Tab.~\ref{tab:ew-lines}. Line ratios used in Sect. \ref{sec:res} and in the discussion of Sect.~\ref{sec:discussion} are reported in Tab. \ref{tab:line-ratios}.
	
	Thanks to the high S/N achieved in the central spatial pixels ($0,~\pm1$), we were able to robustly constrain the contribution of \NIIall (hereafter [\ion{N}{II}]) to the \Halpha emission. Specifically, the inclusion of the [\ion{N}{II}] component is strongly favored, with a difference in the Bayesian Information Criterion of $\Delta(\mathrm{BIC})>10$ when comparing fits with and without [\ion{N}{II}] (see Fig.~\ref{fig:HaNII-fit} in Appendix~\ref{app:NIIcomp}). This level of constraint is not achieved in the $\pm2$ pixels; therefore, all measurements reported for \Halpha in these outer pixels correspond to the blended \Halpha+[\ion{N}{II}] emission.
	
	Given the asymmetric shape of the underlying continuum adjacent to the \lya\ emission line, caused by the presence of the \lya-break blueward of the emission, we fitted the continuum in the rest-frame 1250--1500~\AA\ wavelength range using a linear model implemented with the \texttt{scipy curve\_fit} package. The \lya\ emission line in the stacked galaxy spectra is intrinsically asymmetric, reflecting the dispersion in velocity offset compared to the systemic redshift of the LAEs sample, commonly observed in individual LAEs \citep[e.g.,][]{Verhamme2018, Jung2024, Prieto-Lyon2025}. As a result, a single Gaussian profile does not adequately represent the observed emission of the stack. Therefore, we measured the \lya\ line flux via direct integration over the rest-frame 1200--1240 \AA\ wavelength window. The EW(\lya) was then derived from the measured \lya\ flux and the continuum level extrapolated at the wavelength corresponding to the \lya\ peak.

	\section{Data Analysis and Results}
	\label{sec:res}
	
	Figs.~\ref{fig:fit-spec-pix0}, \ref{fig:fit-spec-pix1} and \ref{fig:fit-spec-pix2} present the observed stacked spectra for each spatial element in the 2D spectrum along with the best-fit models. Emission lines are detected in 5 different spatial pixels, namely pixels 0, +1, -1, +2, -2 in increasing order of distance from the center (i.e., pixel 0). In the following sections, we report the main findings concerning the spatially resolved spectral properties in our stack of LAEs.
	
	Differently from non-LAEs \citep[see e.g.,][]{roberts-borsani2024}, LAEs exhibit a rich set of bright emission lines spanning both the rest-frame UV and optical wavelength ranges. In the UV, \lya\ emission is detected from the central regions out to the outskirts. Among the metal and helium lines, \CIII is robustly detected at all radii, while \OIIIuv+\HeIIL[1640] and \HeIIL are detected only in the central spatial pixels ($0,\pm1$). Notably, also \CIV emission is observed exclusively in the central pixels, suggesting that the physical conditions required to produce high-ionization lines are confined to the innermost regions. 
	
	In the optical regime, \OIIall (hereafter \OIIL[3727]), \NeIIIe, the Balmer series down to \Hdelta, \OIIIL[5007], and \HeIL[5875] are detected from the center to the outer spatial pixels. Additional optical lines, including \HeIL[4471], \SII, and \HeIL[7065], are detected only in the central regions ($0,\pm1$), and [\ion{O}{II}]\textlambda\textlambda 7319,7330 (hereafter \OIIL[7320]) just in the central pixel. Tentative detections of \ion{Ar}{III} are also found in pixels $0$ and $+1$. Other tentative detections are listed in Tab.~\ref{tab:ew-lines}.
	
	We report the results for the EWs of all emission lines in Tab.~\ref{tab:ew-lines}, and we will discuss the implications for the properties of LAEs in Sect.~\ref{sec:lya}. 
	
	\begin{table*}[]
		\centering
		\caption{UV slopes, metallicity and nitrogen-over-oxygen abundance}
		\begin{tabular}{c|ccccc}
			& $\beta_{\rm UV}$  & $\log(Z_{\rm Sanders}/Z_\odot)$ & $\log(Z_{\rm direct}/Z_\odot)$ & $\log(Z_{\rm \textsc{HCm}}/Z_\odot)$ & $\log(\rm N/O)_{\rm \textsc{HCm}}$\\
			\hline

			Pix 2  & $-2.5\pm 0.1$  & $-1.38\pm 0.05$ & $-1.20\pm 0.03$ & $-0.96\pm 0.02$ & N/A\\
			Pix 1  & $-2.07\pm 0.02$  & $-1.20\pm 0.03$ & $-0.85\pm 0.02$ & $-0.85\pm 0.04$ & $-0.3\pm0.2$\\
			Pix 0  & $-1.95\pm 0.02$  & $-1.22\pm 0.02$ & $-0.93 \pm 0.01$ & $-0.91\pm 0.02$ & $-0.5\pm 0.4$\\
			Pix -1 & $-2.32\pm 0.01$ & $-1.23\pm 0.03$ & $-0.80\pm 0.02$ & $-0.90\pm 0.07$ & $-0.8\pm 0.4$\\
			Pix -2 & $-2.29\pm 0.05$  & $-1.26\pm 0.05$ & $-1.04\pm 0.04$ & $-1.00 \pm 0.05$ & N/A\\
			\hline
			\hline
		\end{tabular}
		\tablefoot{Columns: UV continuum $\beta$ slope; metallicity computed from the calibrations reported in \citet{sanders2025}; metallicity computed from the $T_e$-direct method using models from NUVOLOSO project (see Sect. \ref{sec:metal-ion}); metallicity computed from \textsc{HCm}; N/O abundance computed from \textsc{HCm}.}
		\label{tab:uv-dust}
	\end{table*}
	
	\subsection{Photoionization models}
	\label{sec:nuvoloso}
	
	In this section, we briefly describe the photoionization models retrieved from the `detailed analysis of Nebular UltraViolet and Optical emission Lines to Optimize Studies of the iOnized gas' (NUVOLOSO) project (P\'{e}rez-D\'{\i}az et al. in prep), which will be used in the analysis to derive key physical quantities such as gas-phase metallicity, ionization parameter and nitrogen-to-oxygen abundances. 
	
	Photoionization models were computed using \textsc{Cloudy} v25 \citep{gunasekera2025}, where the ionizing source is SSPs accounting for binaries from BPASS v2.3 \citep{byrne2022} characterized by the same metallicity of the gas-phase ISM and accounting for a young burst with age of 1 Myr. Longer ages (up to 3 Myr) induce small differences in the chemical abundance derivation \citep[e.g.,][]{perezmontero2014}.
	
	Model grids have been derived covering a great range of the parameter space. In terms of chemical composition of the gas-phase, the explored metallicities range from 0.008 Z$_{\odot}$ (extremely metal-poor) to 1.8 Z$_{\odot}$ (super-solar) in steps of 0.2 dex. All elements are scaled following the solar proportions \citep{asplund2021}, with the exception of N for which we explore alternative ratios from -2.0 to 0.5 and C which is scaled following log(C/N) = 0.6 \citep{berg2016, steidel2016, perezmontero2017b}. By means of variations of log($U$) which is covered in the range of [-4.0, -1.0], we explore different scenarios accounting for the total number of ionizing photons released by the same SED shape, for fixed H density\footnote{The NUVOLOSO project offers different density profiles, including power-law density models, to explore a wide range of physical conditions. The models used in this work have been computed assuming a constant density profile, since they well reproduce the observed data while minimizing degeneracies. Higher densities are unlikely, since they would imply UV lines to be brighter with respect to the optical counterparts, and this is at odds with the observed data. A power-law density scenario would require an extra free parameter (the power-law index), which increases the degeneracy within the model. Also in this case, UV line fluxes would increase. In general, exploring more complex conditions requires a much larger statistical sample of individual galaxies, and it is beyond the scope of this work.} of 1000 cm$^{-3}$ and assuming the standard dust-to-gas ratio of the Milky Way. These models represent a small subset of the grid of models that constitute the NUVOLOSO project (P\'{e}rez-D\'{\i}az et al. in prep.).

	\subsection{Test of the observed \NII/\Halpha ratio}
	\label{sec:N2Ha-ratio}
	
	A robust determination of the \NII/\Halpha ratio is important for constraining the dust attenuation via Balmer decrements that include \Halpha. To further assess the reliability of the \NII/\Halpha ratios inferred from the best-fitting spectra in pixels $0$ and $\pm1$ (see Sect. \ref{sec:method}), we compared our measurements with predictions from the photoionization models of the NUVOLOSO project (see Sect.~\ref{sec:nuvoloso}; P\'{e}rez-D\'{\i}az et al., in prep.). 
	
	We measure $\log(\NII/\Halpha)=-1.0\pm0.1$ in pixel $0$, $\log(\NII/\Halpha)=-0.77\pm0.15$ in pixel $+1$, and $\log(\NII/\Halpha)=-1.0^{+0.2}_{-0.3}$ in pixel $-1$ (see also Tab.~\ref{tab:line-ratios}). As shown in Fig.~\ref{fig:N2Ha-ratio}, where \NII/\Halpha is compared with \NeIIIe$_{\rm blend}$/\OII, our measurements are in good agreement with model predictions at gas density $\log(n/\rm cm^{-3})=3$, supporting the robustness of the inferred \NII/\Halpha ratios despite the limited spectral resolution. Our results also match \Halpha/(\Halpha+\NIIall)$=0.96$ found in a stack of medium resolution grating spectra of $z>4$ galaxies \citep{sandles2024}, and the \NII/\Halpha ratios found in a sample of 588 galaxies at $1.5<z<7$ using R$=1000$ grating spectra \citep[see Fig. 3 in][]{cameron2026}. As an additional consistency check, we derived the \NII/\Halpha ratio from the $R=1000$ grating spectra of a subset of our parent sample. In particular, for the 53 LAEs in JADES DR4 with available high-resolution spectroscopy \citep{curtislake2025,scholtz2025}, we measure an average value of $\log(\NII/\Halpha) = -1.3 \pm 0.5$. This result is consistent with the value inferred from our stacked spectrum within the uncertainties. The small offset between the two measurements may reflect differences in sample properties, as the JADES subsample is, on average, brighter in $M_{\rm UV}$ and characterized by lower EW(\lya) compared to the full stacked sample.
	
	From the comparison with the models in Fig.~\ref{fig:N2Ha-ratio}, we also infer ionization parameters in the range $-2.0 < \log({\rm U}) < -1.5$ and nitrogen-to-oxygen abundance ratios spanning $-0.5 < \log({\rm N/O}) < 0.0$. As an independent check, we also estimate the N/O abundance in pixel $0$, where the S/N is highest, using the empirical correlation between $\log({\rm N/O})$ and the \NII/\SII line ratio reported by \citet{perezmontero2009} and by \citet{cataldi2025}. This approach yields $\log({\rm N/O})=-0.3\pm0.2$ and $=-0.5\pm 0.1$, respectively, which is consistent with the values inferred from the photoionization models within the uncertainties.
	
	The implications of these constraints for the gas-phase metallicity and nitrogen enrichment are discussed in detail in Sects.~\ref{sec:metal-ion} and \ref{sec:Nenrich}, respectively.

	\subsection{UV slopes and dust correction}
	\label{sec:dust-corr}
	
	The blue UV continuum slopes, $\beta_{\rm UV}$, reported in Tab.~\ref{tab:uv-dust} are derived from the best-fitting spectral models shown in Figs.~\ref{fig:fit-spec-pix0}, \ref{fig:fit-spec-pix1}, \ref{fig:fit-spec-pix2}, assuming $F_\lambda\propto \lambda^{\beta_{\rm UV}}$, and are measured within the interval 1300–2600~\AA\ avoiding bright emission lines. This is approximately the standard Calzetti wavelength range \citep{Calzetti1994}, usually adopted to measure $\beta_{\rm UV}$.
	
	A robust estimate of the dust attenuation, $A_V$, is required to reliably correct emission-line fluxes separated widely in wavelength. However, the low spectral resolution of NIRSpec/prism makes the determination of $A_V$ especially challenging in our data. In particular, the Balmer decrement \Halpha/\Hbeta is affected by blending between \Halpha and \NII; the \Hbeta/\Hgamma ratio critically depends on an accurate measurement of the \OIIIL[4363] line, which is blended with \Hgamma; while the \Hbeta/\Hdelta decrement is sensitive to the signal-to-noise ratio of the underlying continuum, which becomes progressively weaker and noisier in the outer spatial pixels. Indeed, the \Hbeta/\Hdelta ratio is only measured up to pixels $\pm1$ with S/N$>4$. 
	
	Assuming that the \NII contribution to \Halpha inferred from our best-fit models in pixels $0$ and $\pm1$ is reliable (see Sect.~\ref{sec:method} and Sect.~\ref{sec:N2Ha-ratio}), the resulting negative \Halpha/\Hbeta Balmer decrements are consistent with no dust attenuation at all radii. Balmer decrements below Case~B are found even in high S/N individual spectra \citep[e.g.,][]{reddy2015}, supporting no dust contribution. Moreover, deviations from Case~B theory are expected from radiative transfer effects either in very optically thick scenarios \citep{scarlata2024}, or in a density-bounded regime \citep{mcclymont2025}. This scenario is in good agreement with the uniformly blue UV slopes observed across the spatial extent of the stacked LAEs. 
	
	From the \Hbeta/\Hdelta ratio, we infer instead higher dust attenuation values which increase radially  ($A_V=0.7\pm 0.4, 0.9\pm 0.5, 2.1\pm 0.7$ in pixels $0,-1,1$), excluding the pixels $2,-2$ due to the low S/N. This trend, however, is at odds with the observed UV continuum slopes, which become progressively bluer toward the outskirts. Given the large uncertainties, the values are still consistent with Case~B with no attenuation, within 2-3 $\sigma$. Therefore, in the following we adopt no dust attenuation in our analysis.

	\subsection{Line blending and contamination correction.}
	\label{sec:contam}
	
	When measuring \NeIII, we explicitly account for flux contributions from \Heta, \Hzeta\ ($\lambda = 3890.17$~\AA), and \HeIL, all of which are blended with \NeIII\ at the spectral resolution of the NIRSpec/prism. We define the total blended flux as $\NeIIIe_{\rm blend} = \NeIII + \Hzeta + \HeIL + \Heta$.
	
	To estimate the contamination from these additional lines, we proceed as follows. For the Balmer lines \Hzeta\ and \Heta, we measure the $\Hdelta/\NeIIIe_{\rm blend}$ flux ratio from the spectral fits, obtaining values in the range $0.3$–$0.4$. Assuming theoretical Balmer line ratios from Case~B recombination, this implies $\Heta/\NeIIIe_{\rm blend} = 0.08$–$0.11$ and $\Hzeta/\NeIIIe_{\rm blend} = 0.12$–$0.17$.
	
	The contamination from \HeIL is estimated in the central pixels ($0$, $\pm1$) using the detected \HeIL[5876] emission line, assuming Case~B recombination (i.e., $\HeIL[5876]/\HeIL = 1.19$). This yields $\HeIL/\NeIIIe_{\rm blend} = 0.17$–$0.20$ (see Tab.~\ref{tab:line-ratios}). For the outer pixels ($\pm2$), where no \ion{He}{I} line is detected or the flux determination is compromised by the noisy continuum, we assume the same $\HeIL/\NeIIIe_{\rm blend}$ ratio as in the inner regions.
	
	Combining all contributions, we obtain $(\Hzeta + \Heta + \HeIL)/\NeIIIe_{\rm blend} = 0.37$–$0.48$, which corresponds to $\NeIII/\NeIIIe_{\rm blend} = 0.52$–$0.63$ (see Tab.~\ref{tab:line-ratios}). Given the uncertainties associated with the assumption of no dust and the inferred \HeIL contribution, we present the line ratios involving \NeIII corrected for blending in Fig. \ref{fig:app-corr-Ne3}. The corrected values should be considered as lower limits on the intrinsic \NeIII flux.
	
	\begin{figure*}
		\centering
		\includegraphics[width=0.9\linewidth]{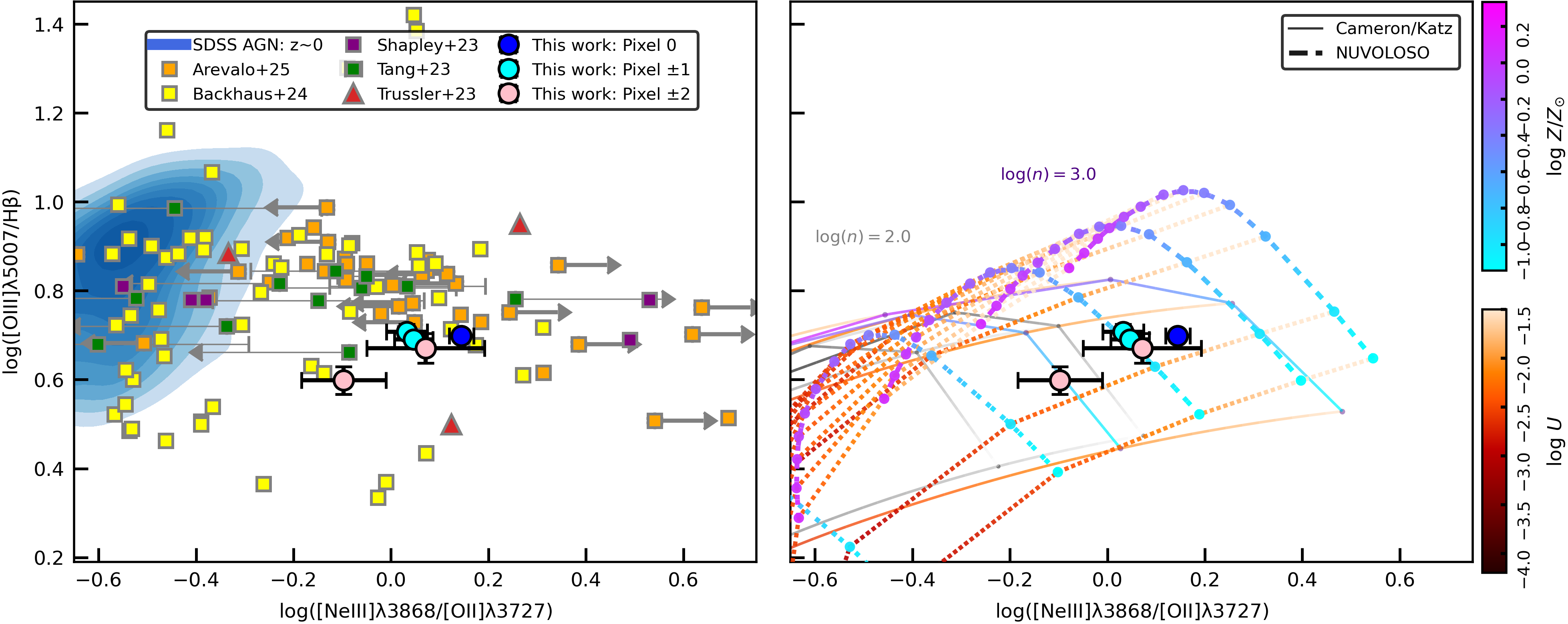}
		\caption{\OIIIL[5007]/\Hbeta--\NeIII/\OIIL[3727] line ratio (`OHNO') diagram for the LAE stack compared with observations and photo-ionization models. Spatially resolved ratios for our stack are plotted as solid circles in both panels and are color  coded based on the distance (in pixel) from the center of the galaxy (i.e., pixel 0).
			Error bars represent standard deviation. (Left) Our LAEs stack (colored dots) is compared with $z\sim 0$ SDSS AGNs as blue colormap with contours, SMACS~06355, 10612 and 04590 \citetext{red triangles; \citealp{trussler2023}; the left-most square of the three is 06355, a type-II AGN identified by \citealp{brinchmann2023}}, high-z \ion{C}{III}] emitters \citep[orange squares,][]{arevalo2025}, a sample of galaxies at $z>2$ \citep[yellow squares,][]{backhaus2024}, a sample of star-forming galaxies in CEERS at $2.7<z<6.5$ \citep[purple squares,][]{shapley2023}, and at $7<z<9$ \citep[green squares,][]{tang2023}. Contours represent the percentiles (from 15\% to 95\% for SDSS with a step of 10\%) of the number of objects in a sample. (Right) overlaid on our results (as in the left panel) are the photoionization models of \citet{cameron2023, katz2023} at hydrogen densities $\log n{\rm [cm^{-3}]}=2.0$ (solid gray scale grid) and $\log n{\rm [cm^{-3}]}=3.0$ (colored solid scale grid), and those from NUVOLOSO at $\log n{\rm [cm^{-3}]}=3.0$ (colored dashed scale grid). The grid shows the variation of the ionization parameter and metallicity (color scales on the right-hand side of the figure, same limits are adopted for gray scale). To guide the eye, lighter colors represent lower metallicity and higher ionization. Y-axis is the same as the left panel.}
		\label{fig:OHNO}
	\end{figure*}

	\section{Spatially resolved properties of Ly$\alpha$ emitters}
	\label{sec:discussion}
	
	\subsection{UV slopes and Dust attenuation}
	\label{sec:disc-uvslope}
	
	Our stacked sample of $z>4$ LAEs exhibits progressively bluer UV continuum slopes $\beta_{\rm UV}$ with increasing radius, indicating a spatial variation in the underlying stellar populations and/or dust content. Our inferred UV slopes ($\beta_{\rm UV}\sim -2.2$ on average) are in good agreement with measurements for single LAEs at $z>5$ \citep{witstok2024,witstok2025}, for a stack of 42 LAEs in JADES \citep{kumari2024}, but also for star-forming galaxies at $z \sim 5$–7 and in the same $M_{\rm UV}$ range reported in the literature \citep{wilkins2011, finkelstein2012, bouwens2014, nanayakkara2024, cullen2023,dottorini2025}. When compared with the stacked LAEs presented in \citet{roberts-borsani2024}, we find that the central UV slope in their stack is steeper than ours in the central pixel by a factor of $\sim1.4$ (i.e., bluer), with $\beta_{\rm UV}=-2.59\pm0.05$, while the agreement improves toward larger radii. However, we note that (i) \citet{roberts-borsani2024} measured $\beta_{\rm UV}$ in a much redder interval (1600-2800 \AA), intrinsically implying bluer slopes, also compared to other works \citep{dottorini2025}. Indeed, \citet{dottorini2025} demonstrated that the choice of the lower limit of the fitted wavelength range impacts the determination of the $\beta$ slope, due to the presence of the Ly$\alpha$ damping wing and additional effects such as deviations from a power-law continuum. (ii) their UV slope is averaged over the spectral extraction window, which is 3-4 pixels; (iii) our parent sample includes approximately five times more LAEs than that of \citet{roberts-borsani2024}, which may partially explain the differences observed in the central regions.
	
	The observed radial trend in $\beta_{\rm UV}$ suggests little to no dust contribution on the largest spatial scales probed by our stacks. This is consistent with our results from the Balmer decrement $\Halpha/\Hbeta$. Indeed, blue UV slopes are found to be connected  with the overall low dust content  also in other single LAEs at similar redshifts \citep[e.g.,][]{Kornei2010, Ono2010, Napolitano2023}, as well as with the weak dust attenuation reported by \citet{roberts-borsani2024} in LAE stacks and by \citet{witstok2024,witstok2025} in single LAEs.
	Alternatively, the trend may also reflect an increasing contribution from younger and/or more metal-poor stellar populations in the outskirts, highlighting the degeneracy between dust and stellar population effects when relying solely on UV continuum slopes. However, as discussed in Sect.~\ref{sec:metal-ion} and Sect.~\ref{sec:lya}, these two scenarios are currently unlikely. While a robust detection of a negative metallicity gradient would support such interpretations, the gradient observed in our data remains tentative and not statistically significant. Moreover, the declining radial trend of EW(\Hbeta) instead points toward centrally concentrated, younger stellar populations, consistent with enhanced recent star formation in the inner regions.
	
	Exploring more complex scenarios, in a non-uniform dust-star geometry, Balmer decrement may be biased towards sightlines with lower obscuration \citep[as indeed suggested by comparing Balmer and Paschen decrements;][]{reddy2026a, reddy2026b}, which implies dust attenuation to be higher than what is inferred solely from the Balmer decrement. In addition, the nebular continuum also contributes to the observed UV continuum slopes \citep{topping2024, cameron2024, saxena2026, katz2025}, and could become more dominant towards the central region if, for instance, the gas density is higher. This scenario is also consistent with the moderate LyC escape found in our sample (see Sect.~\ref{sec:Lyc-escape}).
	
	\subsection{Metallicity and ionization from optical diagnostics}
	\label{sec:metal-ion}
	
	A powerful diagnostic to break the degeneracy between gas-phase metallicity and ionization parameter is the combined use of the \OIII/\Hbeta and \NeIII/\OIIL[3727] line ratios, commonly referred to as the OHNO diagram \citep{backhaus2022}. The left panel of Fig.~\ref{fig:OHNO} shows the OHNO diagram for the observed ratios in our LAE stacks, where we adopt \NeIIIe$_{\rm blend}$/\OII, and compare them with other $z>2$ galaxy samples \citep[see figure caption for details,][]{trussler2023, arevalo2025, shapley2023,backhaus2024,tang2023}, the population of $z \sim 0$ SDSS AGNs (left panel), and photoionization models \citep[right panel,][]{cameron2023, katz2023}. In the left panel, our measurements are broadly consistent with the high-redshift galaxy population, yet they lie toward the lower end of the \OIII/\Hbeta distribution.
	
	The right panel of Fig.~\ref{fig:OHNO} presents a comparison with photoionization models from \citet{cameron2023, katz2023}. In particular, models with gas density $\log(n/\rm cm^{-3})=3$ provide the best overall agreement with the observed line ratios. At fixed metallicity, lower-density models that reproduce our data require ionization parameters that are unusually high for galaxies in this redshift range \citep[i.e. $\log(U)>-1$,][]{sanders2023b, cleri2025, cleri2026}. Visually comparing our best-fitting values with the models, we infer an average metallicity of $\log(Z/Z_\odot)\sim -0.8\pm0.1$ (i.e., $12+\log(\rm O/H)=7.89\pm0.10$) across all pixels, and a tentative gradient in ionization parameter, having $\log(U)\simeq -1.5$ in the central regions, decreasing slightly to $\log(U)\simeq -1.75$ in pixel $-2$. When adopting the corrected \NeIII/\OIIL[3727] ratios (lighter symbols in Fig.~\ref{fig:app-corr-Ne3}; see also Sect.~\ref{sec:contam}), the inferred metallicity increases by $\sim0.2$ dex, while the ionization parameter decreases by a comparable amount. These values for the ionization parameter are also in agreement with those inferred from the comparison of the \NII/\Halpha vs \NeIII$_{\rm blend}$/\OIIL[3727] diagnostic with the NUVOLOSO models.
	
	To improve the level of accuracy in constraining both metallicity and ionization parameter, we further explore two independent methods: the $T_e$-direct method using NUVOLOSO models and \textsc{HII-CHI-Mistry} photo-ionization models.
	
	Owing to the high S/N achieved in the stacked spectra at all radii for \OIII, \Hbeta, and the blended \Hgamma+\OIIIL[4363] feature, we were able to derive gas-phase metallicities using the direct $T_e$ method and relying on predictions from the NUVOLOSO grid of models to account for the contribution of all blended lines. Assuming negligible dust attenuation, the ratio between \OIIIL[4959, 5007] and \OIIIL[4363]+\Hgamma in comparison to model predictions (see P\'{e}rez-D\'{\i}az et al. in prep.) yields electron temperatures of $T_e(\OIII)=15{,}200\pm500$~K, $14{,}400\pm800$~K, $13{,}600\pm1{,}000$~K, $18{,}500\pm1{,}200$~K, and $16{,}500\pm1{,}300$~K in pixels $0$, $+1$, $-1$, $+2$, and $-2$, respectively. As a further test, we verified that these values are in agreement within uncertainties with $T_e(\OIII)$ derived from the observed \OIIIL[4363]/\OIIIL, assuming that [\ion{O}{III}]$\lambda$4363$=$[\ion{O}{III}]$\lambda$4363+\Hgamma-(\Hbeta$\times$0.47) for Case~B with no dust. Specifically, by means of \texttt{Pyneb}, we have $T_e(\OIII)=18{,}000\pm2{,}000$~K, $16{,}000\pm4{,}000$~K, $12{,}000\pm4{,}000$~K, $27{,}000\pm10{,}000$~K, and $21{,}000\pm10{,}000$~K in pixels $0$, $+1$, $-1$, $+2$, and $-2$. 
	
	Adopting the results from NUVOLOSO, the electron temperature associated with the O$^{+}$ zone is estimated following \citet{campbell1986} as $T_e(\OII)=0.7 \times T_e(\OIII)+3000~\rm K$. Ionic and total oxygen abundances are then computed using \texttt{pyneb}, assuming that oxygen resides entirely in the $\rm O^{2+}$ and $\rm O^{+}$ states within \ion{H}{II} regions. The predictions from our photoionization models reveal some amounts of neutral O and highly ionized O$^{3+}$ fractions, which add up to 0.13 dex to the derived metallicity from  $\rm O^{2+}$ and $\rm O^{+}$. Moreover, given the inferred $T_e(\OII)\sim 14{,}000$~K and the detection of both \OIIL[3727] and \OIIL[7320] in the central pixel, we estimated the electron density to be of the order $\log(n_e/\rm cm^{-3})=3.4-4.0$, accounting for uncertainties on the line ratio and the electron temperature, which is an upper limit on $n_e$ given the assumption of no dust. This result  further supports the evidence of high electron density inferred from the comparison with photoionization models in the OHNO diagram. Although the model grids do not reach values higher than $\log(n_e/\rm cm^{-3})=3.0$, a higher electron density would imply systematically lower values of $\log(U)$, leaving the metallicity estimates and the trends mostly unaffected.
	
	Based on the observed \OIII/\Hbeta and \OII/\Hbeta ratios, we find a clear decrease in metallicity (since O dominates the metal budget) with galactocentric radius, from $12+\log(\rm O/H)=7.76$–$7.84$ in the central regions (pixels 0,$\pm 1$) to $12+\log(\rm O/H)=7.49$–$7.69$ in the outskirts (see Tab.~\ref{tab:uv-dust}). We also estimate metallicities using the OHNO-based diagnostic from our photoionization models, explicitly accounting for the blending of \NeIII with \Hzeta, \Heta, and \HeIL, and recover the same radial trend as obtained with the $T_e$-based method (see Tab.~\ref{tab:uv-dust}). 
	
	\begin{figure}
		\centering
		\includegraphics[width=1\linewidth]{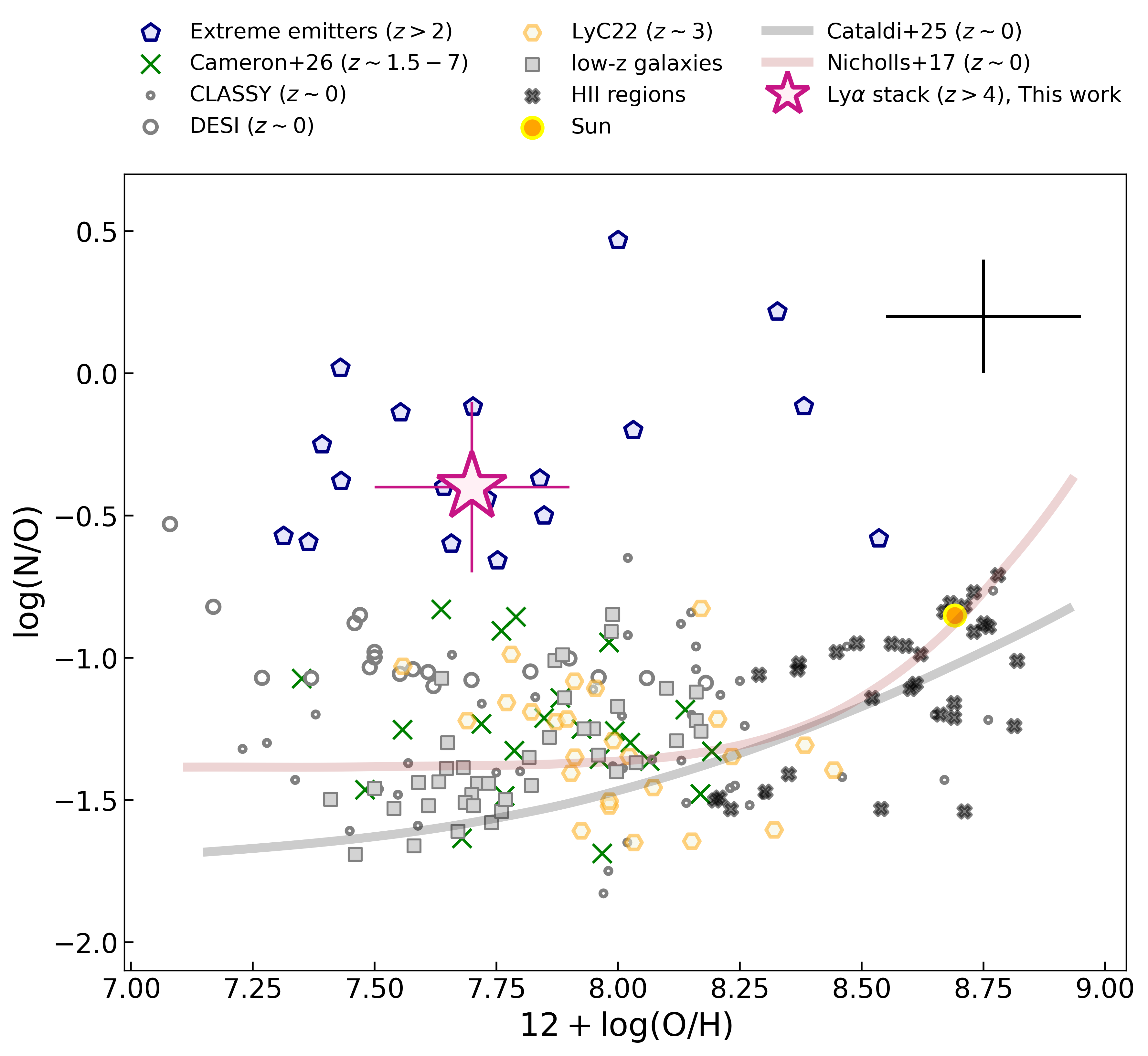}
		\caption{Nitrogen-over-oxygen abundance vs metallicity in terms of $12+\log(\rm O/H)$. Our result is shown as a magenta star, while results from literature are coded as in the legend \citep{schaerer2024, topping2024,castellano2024, Napolitano2025b, zavala2024,izotov2023,cameron2026,bhattacharya2025,arellano2025,isobe2023,naidu2026,martinez2025,marques2024}. We also show the relation derived in \citet{cataldi2025} as a brown line and in \citet{nicholls2017} as a gray line. The black cross in the upper right corner represents the average error on metallicity and N/O for the literature compilation.}
		\label{fig:Nemitters}
	\end{figure}

	\begin{figure*}
		\centering
		\includegraphics[width=1\linewidth]{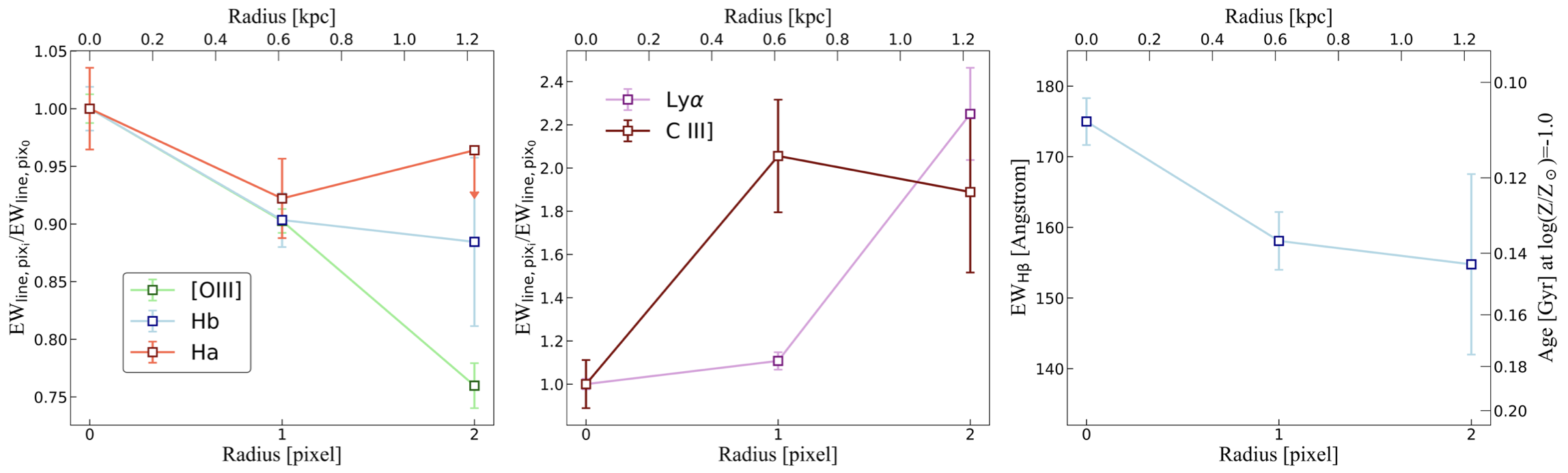}
		\caption{Averaged radial profile of the rest-frame EWs of the brightest emission lines. (Left) The panel shows the comparison between the radial profiles of the EWs of \OIII, \Hbeta and \Halpha normalized to their EW at pixel 0. The unconstrained contribution of [\ion{N}{II}] to the \Halpha flux yields upper limits for the EW(\Halpha) in pixels $\pm2$. (Center) Averaged radial profile for EW(\lya) and EW(\CIII). (Right) Averaged radial profile for EW(\Hbeta). The right y-axis displays values of age of the galaxy at fixed metallicity.}
		\label{fig:EW-comp}
	\end{figure*}
	
	\begin{figure}
		\centering
		\includegraphics[width=0.9\linewidth]{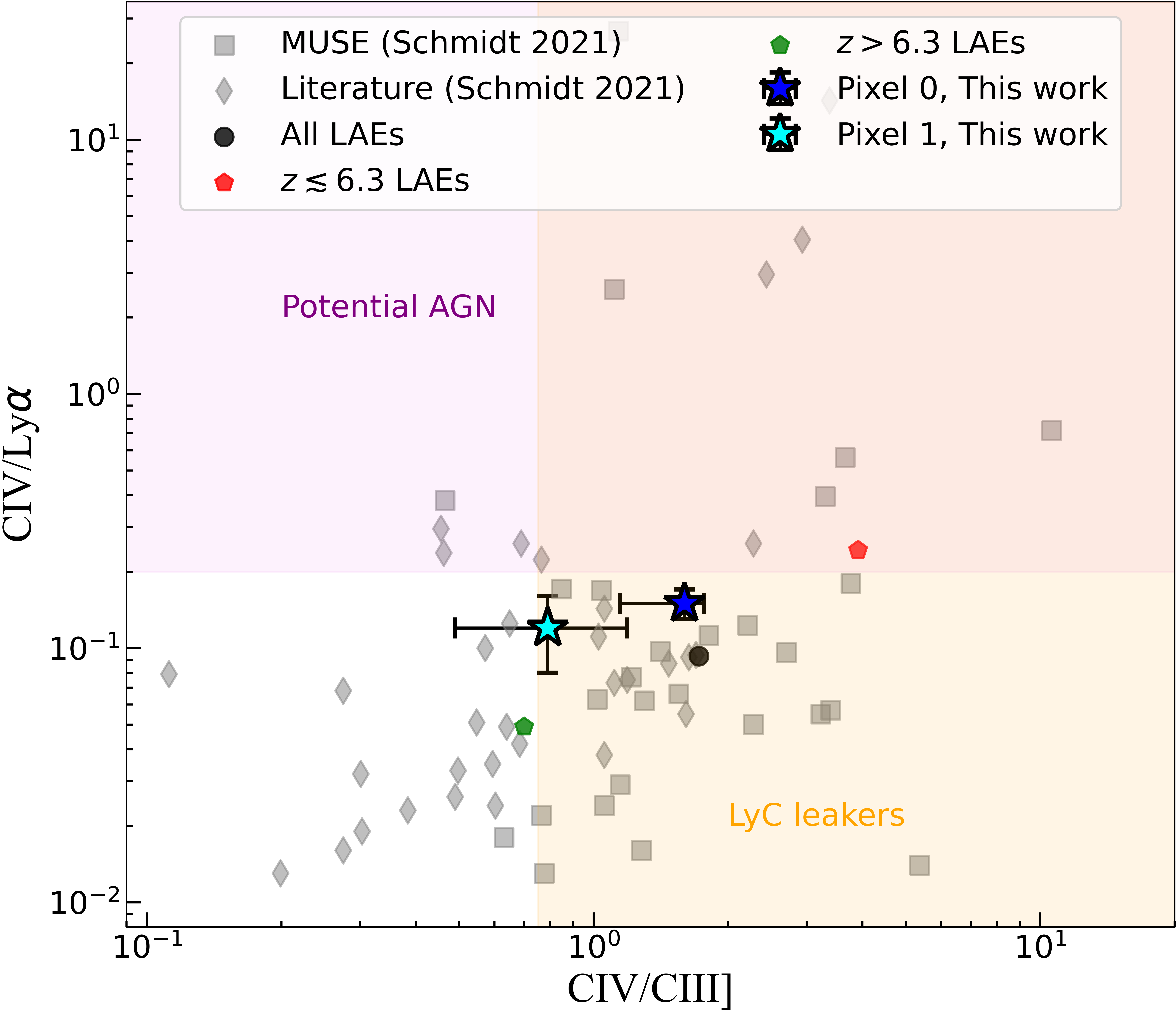}
		\caption{\CIV/\lya ~vs \CIV/\CIII. Our result for pixel 0 (1) is shown as a blue (cyan) star. We show for comparison other LAEs stacks \citep[red, green hexagons and black circle,][]{kumari2024} and a sample of galaxies at $1.5<z<6.4$ \citep{schmidt2021}. The yellow region defined by \CIV/\CIII$>0.75$ identifies possible LyC leakers \citep{schaerer2022}, while the violet one with \CIV/\lya$>0.2$ identifies possible AGN candidates \citep{kumari2024}.}
		\label{fig:Cdiagn}
	\end{figure}
	
	Another approach to estimate chemical abundances, relying on predictions from photoionization models, is using \textsc{HII-CHI-Mistry}\footnote{All versions of HII-CHI-Mistry are publicly available at \url{https://home.iaa.csic.es/~epm/HII-CHI-mistry.html}.} \citep[\textsc{HCm}][]{perezmontero2014, perezmontero2019, perezmontero2020}. In short, \textsc{HCm} estimates chemical abundances (namely 12+log(O/H) and log(N/O), and the ionizing conditions, log($U$)) from the predictions of a large set of photoionization models that account for the diversity of physical and chemical conditions in galaxies. We employed the optical version \textsc{HCm} v5.5 and, given the nature of the galaxies that constitute our sample, we relied on the predictions from \textsc{BPASS} models, assuming empirical relations derived from Extreme Emission Line Galaxies \citep[EELGs][]{perezmontero2021}. Under the assumption of no dust attenuation, the code yields 12+log(O/H) = $7.7-7.8$ across pixels (see Tab.~\ref{tab:uv-dust}), which is in agreement with our previous estimates. Additionally, \textsc{HCm} independently determines the N/O abundance. In our case, we found $-0.8<\log(\rm N/O)<-0.3$ (see Tab.~\ref{tab:uv-dust}) indicating high N/O in LAEs in agreement with the values reported in Sect.~\ref{sec:N2Ha-ratio}.  Similarly high N/O is found in galaxies that also show emission of \ion{N}{IV}] or \ion{N}{III}] \citep[see e.g.,][more complete refs. in Sect.~\ref{sec:Nenrich}]{isobe2025, zhu2026}. However, these lines are undetected in our stack. Based on the derived ionization parameter, metallicity, N/O abundance and the \CIII detection in our stack, we used NUVOLOSO to estimate the expected \CIII/\ion{N}{III}] ratio, obtaining \CIII/\ion{N}{III}]=15. This makes the \ion{N}{III}] undetectable given the sensitivity of the stack, and also implies the non-detectability of \ion{N}{IV}]. A more detailed discussion of the N/O enhancement is reported in Sect.~\ref{sec:Nenrich}.
	
	Finally, if deriving the metallicity from \citet{sanders2025} calibrations using the \OIII/\Hbeta ratio, we find values of $12+\log(\rm O/H)$ that are systematically lower by $\sim 0.3$ dex than those inferred above from other diagnostics (see Tab~\ref{tab:uv-dust}). This discrepancy can be partially explained by the fact that Sanders' calibrations do not account for the presence of states higher than $\rm O^{2+}$, while the robust detection of \HeIIL[4686] points to the need to account for higher ionized species \citep{perezmontero2017, perezdiaz2022}.
	
	In summary, the metallicities derived using the $T_e$-based method and \textsc{HCm} are broadly consistent with each other and with those inferred from the photoionization models of \citet{cameron2023, katz2023}. The small offsets observed, at the level of $\sim0.1$–$0.2$ dex, can be attributed to differences in model assumptions, including variations in the adopted stopping criteria for the photoionization calculations. Overall, we find that LAEs are characterized by sub-solar gas-phase metallicities, typically at the level of $\sim10$–15\% of the solar value. These results are in agreement with the general population of galaxies at $z>3$ \citep{curti2024,venturi2024,hu2024}, but smaller values of metallicity are reported in other smaller LAE stacks, with $Z\sim 6\% Z_\odot$ \citep{kumari2024,roberts-borsani2024}. However, \citet{roberts-borsani2024} inferred metallicity from Sanders' calibrations, which could bias the results towards lower values (see discussion above).
	
	We also found evidence for either negative or flat metallicity gradient in LAEs, in which the metallicity decreases toward the outskirts (see Tab.~\ref{tab:uv-dust}). Negative gradients are thought to arise typically from an inside-out galaxy formation scenario, where stars form earlier in the inner parts of the galaxy, which turn out to be more chemically enriched than the outskirts. Conversely, flatter trends might arise in the case of mergers or outflows that mix and redistribute the gas in the galaxy.
	Metallicity gradients have been extensively studied at high redshift in the past few years: a large observational effort led by \citet{li2025} found that at $z>5$ galaxy centers are more metal rich, exhibiting negative metallicity gradients of $\sim -0.4$ dex kpc$^{-1}$. This was interpreted as an initial phase of galaxy growth dominated by inside-out mode, with inefficient feedback and gas mixing. Other studies carried out with  NIRSpec/IFU seem to point to flatter gradients (although with large scatter), such as those found in a few interacting galaxies \citep{carniani2024}, or those studied by \citet{fujimoto2025} in the ALPINE-CRISTAL sample. However, most of these studies focused on galaxies with stellar masses larger than the bulk of our LAEs sample. 
	
	\subsection{Enhanced N/O in LAEs}
	\label{sec:Nenrich}
	
	In Fig.~\ref{fig:Nemitters}, we compare the N/O abundance vs metallicity of our LAE stack with a compilation of nitrogen emitters spanning a wide range of redshifts \citep{schaerer2024, topping2024,castellano2024, Napolitano2025b, zavala2024,izotov2023,cameron2026,bhattacharya2025,arellano2025,isobe2023,naidu2026,martinez2025,marques2024}. Interestingly, high N/O ratio has also been found in Haro 11, a system with both measured LyC and \lya\ escape \citep{James2013, komarova2024}. The average error on metallicity and N/O for the literature compilation is reported as black cross in the upper right corner of the plot. Our result is shown as a magenta star that conservatively accounts for uncertainties arising from the metallicity determination using different models, the spatial variation of metallicity within the stack (yielding $12+\log(\rm O/H)= 7.7\pm 0.2$, see Sect.~\ref{sec:metal-ion}), and from the N/O determination using different relations and at different radii (yielding $\log(\rm N/O)=-0.4\pm 0.3$, see Sects.~\ref{fig:N2Ha-ratio}, \ref{sec:metal-ion}). We find that LAEs exhibit enhanced nitrogen-over-oxygen abundance, which is comparable to that observed in other high-redshift star-forming galaxies. Although the available data prevent us from ascertaining the origin of the observed high N/O ratio, we explore two possible scenarios.
	
	In recent years, mounting evidence has pointed to a growing population of nitrogen-enriched galaxies at high redshift, including both star-forming systems and AGN hosts \citep{schaerer2026, cataldi2025, curti2024, Napolitano2025b, flury2025, isobe2025, tripodi2025, ji2024, rizzuti2025, stiavelli2025, topping2024, zhu2025, zhu2026, hayes2025}. In particular, \citet{cataldi2025} presented a comprehensive analysis of $\sim500$ nitrogen emitters at $z\sim2$–6, showing that high-redshift galaxies exhibit a systematic enhancement in N/O relative to the local relation, with $\Delta({\rm N/O}) \sim 0.18$ dex on average. This enhancement is metallicity dependent, reaching $\sim0.3$–0.4 dex at $12+\log(\rm O/H)\lesssim8.1$. This turn-over point is, by definition, affected as well by the turn-over point from the N/O vs O/H relation, which changes slightly depending on the methodology used for the chemical abundance estimations \citep{vincenzo2016, andrews2013, perezdiaz2021}. A similar picture emerges for AGN host galaxies: a stacking analysis of Type~I and Type~II AGN hosts at $z\sim4$–7 reports elevated nitrogen abundances, with $\log(\rm N/O)>-0.6$ \citep{isobe2025}. Nitrogen enhancement has been explained invoking several processes. For instance, it has been suggested that super star clusters containing Wolf-Rayet stars and massive stars ($M\gtrsim 10^4~\rm M_\odot$) are the most likely culprits of N-enhancement \citep{berg2026b, zhu2026, marqueschaves2024, charbonnel2023}. In detail, massive stars with low metallicity and high angular velocity may produce nitrogen in the first Myrs of their existence: the high rotation favors the transportation of the C into the core causing a bottleneck in the CNO cycle. This makes N destruction highly inefficient and, therefore, nitrogen accumulates and is easily ejected in the surrounding medium  through stellar winds or stellar collapse \citep{meynet2002, zhu2026}. Interestingly, based on this scenario, some high-z `N-enhanced' galaxies have also been interpreted as proto-globular clusters that would later evolve into globular clusters in the local universe \citep{cameron2023, senchyna2024, ji2025, schaerer2025}. Alternatively, the nitrogen enrichment may arise from pollution from Pop III stars \citep{Maiolino2024b}, AGB stars, and tidal disruption of stars from BH encounters \citep{cameron2023, johnson2023, watanabe2024}. Specifically, the presence of Pop III candidates in the environment of GN-z11 suggests that nearly pristine star formation may coexist with more evolved systems, potentially imprinting unusual CNO abundance patterns on nearby gas \citep{Maiolino2024b}. A more classical channel involves intermediate-mass AGB stars \citep{dantona2023}; however, at the highest redshifts this scenario requires very early star formation, and rapid stellar evolution. Finally, if a central black hole is present, tidal stripping or disruption of stars could expose and inject CNO-processed, nitrogen-rich material into the nuclear gas, naturally producing a highly localized enhancement in N/O and N/C \citep{cameron2023, johnson2023}. These scenarios are therefore appealing because they can operate on small spatial scales, but they generally require fine-tuned conditions and remain difficult to distinguish observationally.

	On the other hand, LAEs may display elevated N/O ratios due to oxygen depletion rather than nitrogen enrichment. Evidence of oxygen depletion has been found in a sample of infrared galaxies by \citet{perezdiaz2024}. Thanks to constraints on stellar mass, metallicity, star formation rate and N/O they were able to single out the origin on N/O enhancement finding that, for some galaxies, episodes of pristine gas infall due to mergers could have reduced the amount of oxygen, directly boosting the N/O ratio. Other works favor the interpretation of oxygen depletion caused by either supernova winds, dilution by pristine gas, or a combination of high star-formation and differential galactic winds \citep{stiavelli2025, mcclymont2026,rizzuti2025,arroyo2023}. To assess whether the observed enhancement in N/O could be entirely driven by oxygen depletion, we computed the values of $\log({\rm N/O})$ and $12+\log({\rm O/H})$ expected on the local relation under this assumption. Starting from the observed gas-phase metallicity ($12+\log({\rm O/H})=7.7$), and the enhanced abundance ratio ($\log({\rm N/O})=-0.4$), we find that the corresponding point on the local relation would have $\log({\rm N/O})=-1.28$ and $12+\log({\rm O/H})=8.59$. This would require, on average, about seven times more oxygen than observed in the gas phase, corresponding to an oxygen depletion fraction of $\sim 87\%$ for the stacked LAE sample. Such a high level of oxygen depletion has not been observed in either local or high-redshift galaxies, nor is it predicted by standard depletion models. Indeed, adopting a standard depletion scenario \citep{jenkins2009}, we would expect a maximum oxygen depletion of about 40\% on average, which would shift $\log({\rm N/O})$ downward and $12+\log({\rm O/H})$ upward by only $\sim 0.3$ dex. Therefore, while we cannot exclude some contribution from oxygen depletion to the observed N/O enhancement, a scenario in which the entire offset is produced by oxygen depletion alone appears highly unlikely.

	\begin{figure*}
		\centering
		\includegraphics[width=0.465\linewidth]{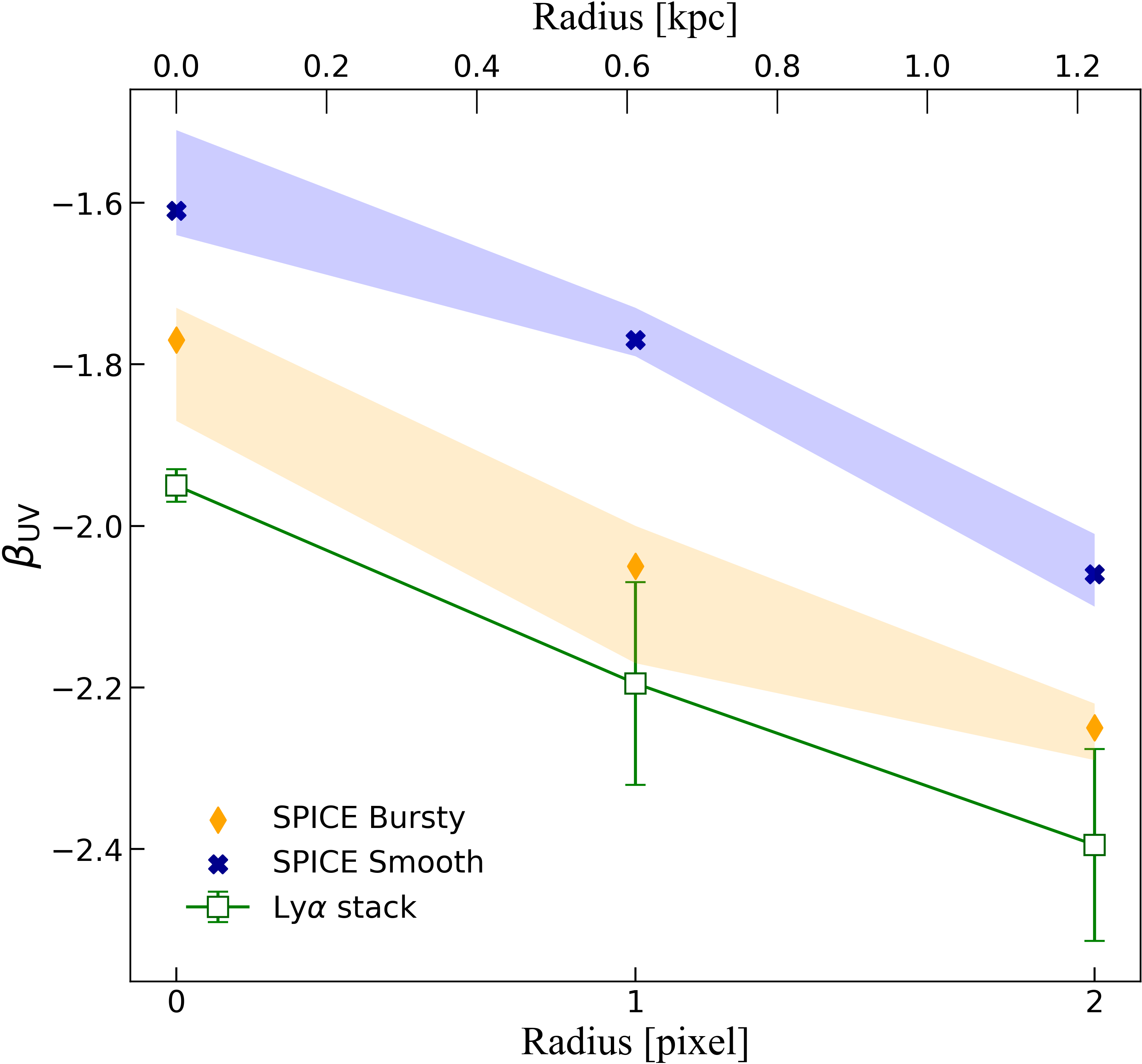}
		\includegraphics[width=0.45\linewidth]{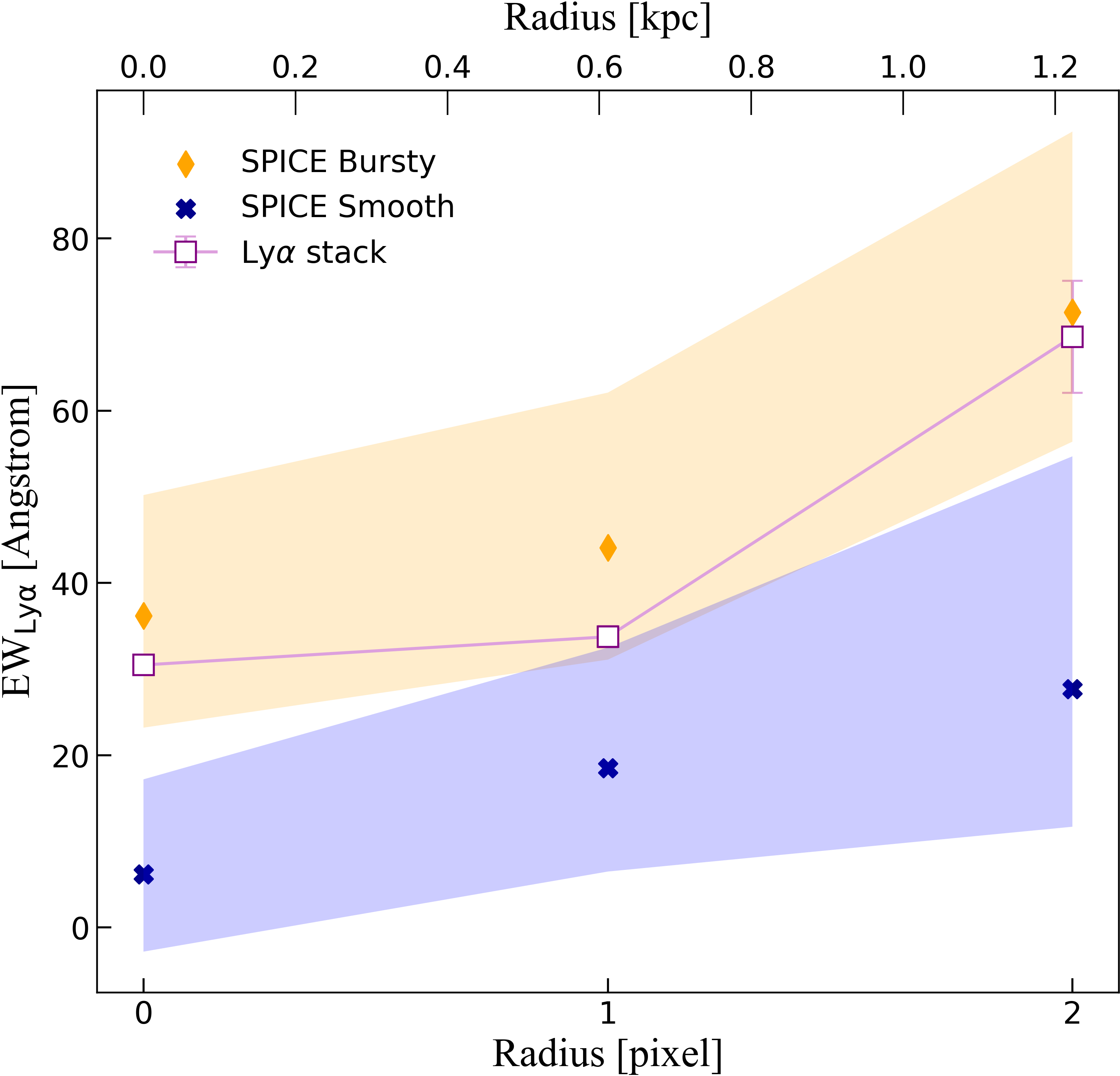}\\[0.1cm]
		\includegraphics[width=0.45\linewidth]{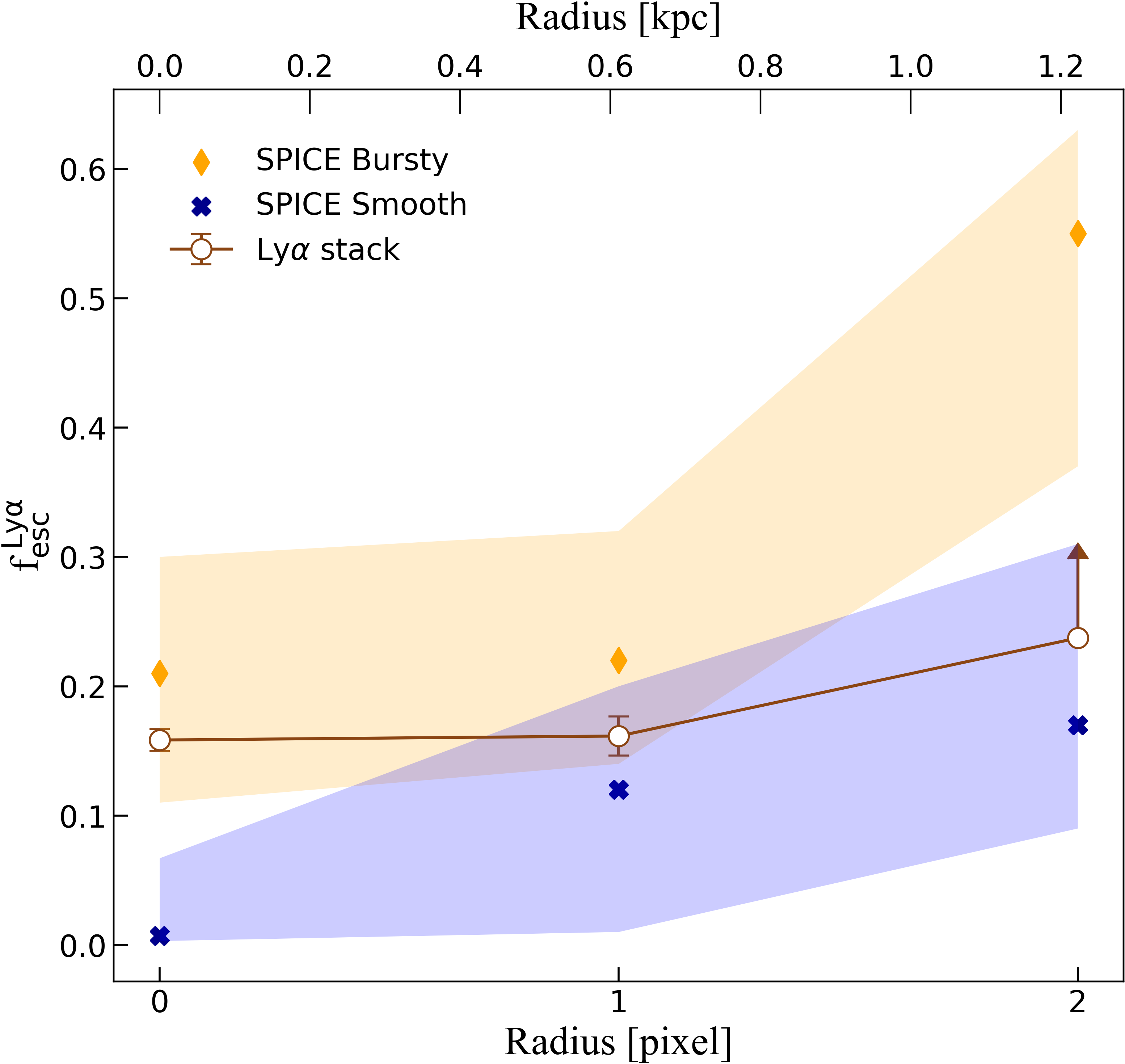}
		\includegraphics[width=0.46\linewidth]{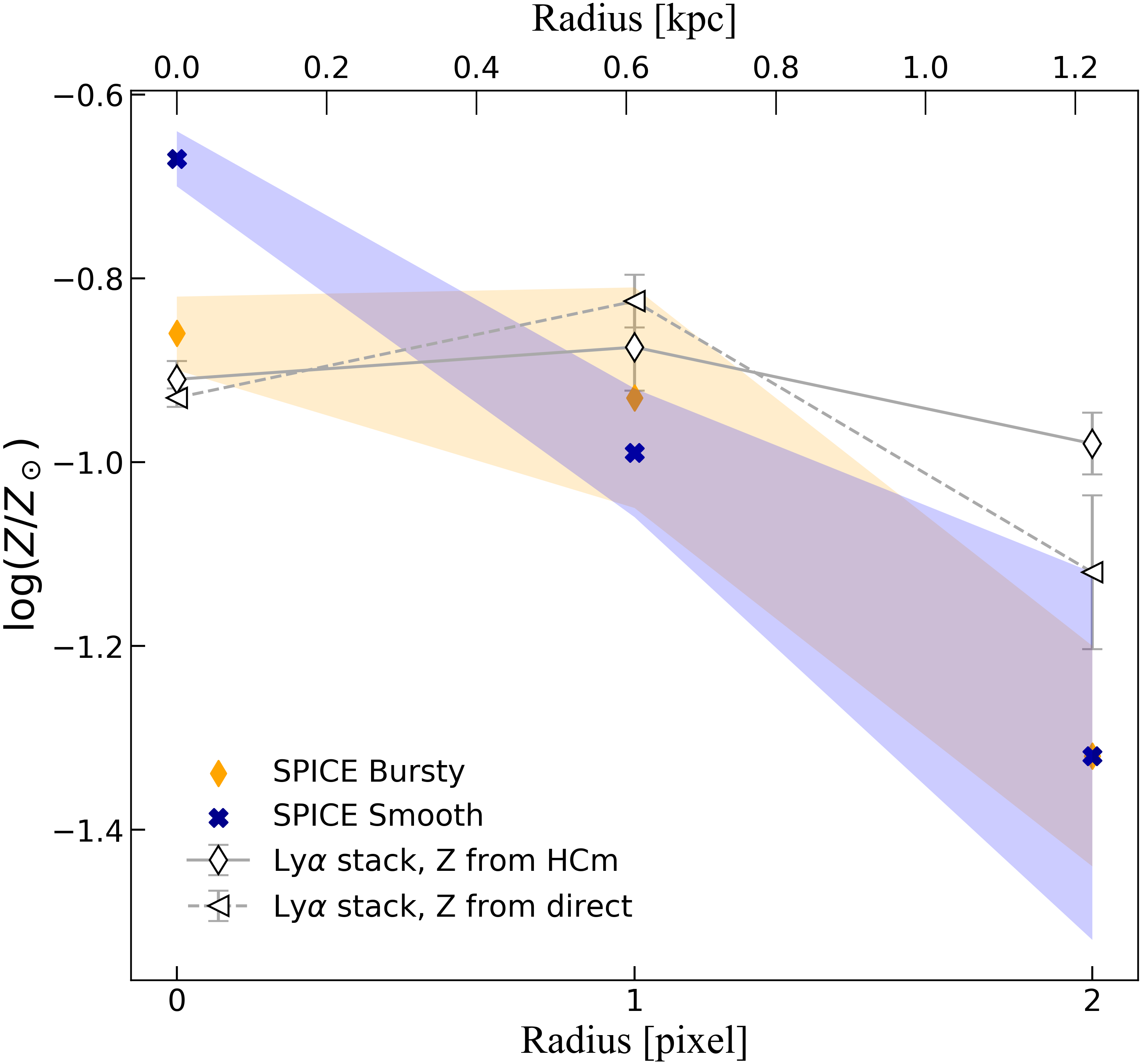}
		\caption{Averaged radial profiles of the observed continuum UV slope (Top left), the rest-frame EW(\lya) (Top right), the escape fraction of \lya, $f_{\rm esc}^{\rm Ly\alpha}$ (Bottom left) and the gas-phase metallicity (Bottom right) compared with SPICE simulations with smooth and bursty stellar models.  
		}
		\label{fig:lya-prop}
	\end{figure*}
	
	\subsection{Ly$\alpha$ properties and connection with UV-optical emission lines}
	\label{sec:lya}
	
	The left and central panels of Fig.~\ref{fig:EW-comp} show the averaged radial profiles of the equivalent widths (EWs) of \Halpha, \Hbeta, \OIII, \CIII, and \lya. We checked that all observed radial trends are physical rather than driven by the convolution of intrinsically different line and continuum profiles with the wavelength-dependent PSF (see Appendix~\ref{sec:app-check-PSF}). While the optical non-resonant emission lines display mild to steeply declining trends with increasing radius, the EW(\lya) exhibits a pronounced increase toward the outskirts. Since the EW traces the ratio between the considered line flux and adjacent continuum, we interpret the spatially increasing trend of EW(\lya) as suggestive that spatially extended \lya\ emission with respect to the compact UV-continuum is due to resonant scattering \citep[e.g.,][]{kusakabe2022, Jung2024, Scholtz2024}.
	Such a trend has also been observed in local \lya\ emitters from the LaCOS survey, which consists of green pea-like galaxies around $z\sim 0.2$ \citep{saldanalopez2026}. They found that the scale length of \lya\ halos can be 10$\times$ larger than that in the UV, for that sample of starbursts.
	
	Interestingly, the EW(\CIII) also increases in pixels $>0$ by a factor of 2, despite the fact that it is not a resonant line. 
	
	In star-forming galaxies, the EWs of Balmer recombination lines such as \Hbeta and \Halpha are known to trace the cosmic evolution of the specific star formation rate (sSFR; \citealt{fumagalli+2012, sobral+2014}). For a fixed stellar population age (see right panel of Fig.~\ref{fig:EW-comp}), EW$(\Hbeta)$ can therefore be used as a rough proxy for sSFR. In this context, the higher EW$(\Hbeta)$ observed in the central regions suggests that our sample is dominated by galaxies experiencing centrally concentrated star formation, leading to the rapid build-up of stellar mass in their inner regions (e.g., \citealt{baker2023, dekel2009, krumholz2018, tacchella2016, tripodi2023, zolotov2015}; see also \citealt{zhang+2012} for a local dwarf galaxy analogue). A qualitatively similar, but even stronger, trend has been reported for stacked AGN host galaxies at $z>4$ \citep{Tripodi2024c}. Although the upper limits on \Halpha in pixels $\pm2$, due to the uncertain contribution of \NII, are consistent with this picture, the trend at these radii remains tentative given the large uncertainties on EW(\Hbeta).
	
	Using our constraints on the \lya\ and \Halpha fluxes, we can estimate the Ly$\alpha$ escape fraction ($f_{\rm esc}^{\rm Ly\alpha}$), defined as the ratio between the observed and intrinsic \lya\ flux. The observed \Halpha flux serves as a proxy for the intrinsic \lya\ emission under the assumption of Case~B recombination and a null dust attenuation, such that
	
	\begin{equation}
		f^{\rm Ly\alpha}_{\rm esc}=\dfrac{F^{\rm Ly\alpha}_{\rm obs}}{F^{\rm Ly\alpha}_{\rm int}}=\dfrac{F^{\rm Ly\alpha}_{\rm obs}}{8.2\times F^{\Halpha}_{\rm obs}}.
	\end{equation}

	Applying this relation at each radius, we find $f^{\rm Ly\alpha}_{\rm esc}\simeq16\%$ in the central regions, increasing to $f^{\rm Ly\alpha}_{\rm esc}\gtrsim24\%$ in the outskirts. The resulting averaged radial trend is shown in the bottom panel of Fig.~\ref{fig:lya-prop}. Moreover, based on the test performed in Appendix~\ref{sec:app-check-PSF}, the value of $f_{\rm esc}^{Ly\alpha}$ derived at pixel 1 may also be considered as an upper limit. Comparably high \lya\ escape fractions have been measured in individual LAEs at $z>6$ \citep{saxena2024, Napolitano2024,llerena2025b}, supporting the idea that our stacked sample is representative of the broader LAE population at $z>4$. Furthermore, the average $f^{\rm Ly\alpha}_{\rm esc}\gtrsim20\%$ found in our stack is in agreement with the value derived from the empirical $f^{\rm Ly\alpha}_{\rm esc}-\mathrm{EW}(Ly\alpha)$ relation reported in \citep{saxena2024} at EW(\lya)=60\AA, which matches the median EW(\lya)$_{\rm median}$ in our analysis. Since both \lya\ and \Halpha\ originate from star-forming regions, the increasing trend of $f^{\rm Ly\alpha}_{\rm esc}$ as a function of the radial profile further supports our previous finding of a more spatially extended \lya\ emission due to its resonant scattering process with respect to the non-resonant Balmer lines.

	Finally, we estimate the ionizing photon production efficiency, $\xi_{\rm ion}$,  in two ways: (1) using the Balmer line fluxes and (2) using the EW(\OIII) as a proxy. Since the stack is normalized to the flux at 1500 \AA, we compute $\xi_{\rm ion}$ directly from the estimated fluxes of \Halpha and \Hbeta in pixel 0, which has the highest S/N, considering that $L({\rm H\alpha})=1.36\times 10^{-12}\rm N(H^0)$, and $L({\rm H\beta})=4.87\times 10^{-13}\rm N(H^0)$ \citep{leitherer1995}. Assuming no dust, case~B recombination and $f_{\rm esc}^{LyC}=0$, we obtain $\log(\xi_{\rm ion})=25.07\pm 0.02$ from \Halpha and $\log(\xi_{\rm ion})=25.12\pm 0.02$ from \Hbeta. These can be considered as lower limits as $f^{\rm LyC}_{\rm esc}>0$ (see Sect. \ref{sec:Lyc-escape}), but we do not compute them explicitly given the high uncertainties on $f^{\rm LyC}_{\rm esc}$ . (2) Adopting the empirical relation calibrated for star-forming galaxies at $z\sim4$–10 by \citet{llerena2025}, averaged over all radii, we find $\log(\xi_{\rm ion}/{\rm Hz~erg^{-1}})\simeq25.2\pm0.1$, where the uncertainty includes systematic errors in the adopted calibration. All the derived values lie close to the mean $\xi_{\rm ion}$ reported in the literature for galaxies with average $M_{\rm UV}\simeq-19$ \citep{llerena2025,pahl2025} and $M_{\rm UV}\simeq-18, ~3<z<8$ \citep{papovich2025,prieto2023,simmonds2024}, in good agreement with the median UV magnitude of our sample, $M_{\rm UV, median}=-18.7$. Moreover, it agrees well with the values of $\xi_{\rm ion}$ inferred by semi-analytical model at $4<z<10$ \citep{yung2020b}.
	
	\subsection{The connection between UV-optical lines and LyC escape}
	\label{sec:Lyc-escape}
	
	Since the escape of \lya\ and LyC photons is facilitated by similar interstellar medium geometries and low column densities \citep{verhamme2017}, \lya\ is widely considered an important tracer of Lyman continuum (LyC) escape. Indeed, several \lya\ properties correlate with LyC escape, including the \lya\ equivalent width, the \lya\ escape fraction, and the \lya\ velocity offset \citep{marchi2018, verhamme2017}.
	
	Empirically, $f_{\rm esc}^{\rm Ly\alpha}\sim 20\%$ typically corresponds to $f_{\rm esc}^{\mathrm{LyC}}\sim 5$--$10\%$, with the latter generally smaller \citep[e.g.,][]{flury2022a}. 
	The LaCOS resolved study of \lya\ emission confirms correlations between \lya\ EW and \lya\ luminosity with $f_{\rm esc}^{\mathrm{LyC}}$ \citep{lereste2025}, although they also find an anticorrelation between \lya\ halo size and $f_{\rm esc}^{\mathrm{LyC}}$ \citep{saldanalopez2026}. But overall, our inferred \lya\ escape fractions are compatible with moderate LyC leakage.
	Another proposed tracer of LyC escape is the \OIIIall/\OIIall ratio (O32; \citealt{jaskot2013, izotov2018}). Although recent studies indicate that the correlation between O32 and $f_{\rm esc}^{\mathrm{LyC}}$ is not tight—likely due to complex gas geometry and kinematics—a high O32 appears to be a necessary (though not sufficient) condition for significant leakage \citep[e.g., as shown for Haro 11 by][]{keenan2017}. For instance, \citet{flury2022a} showed that among galaxies with $\log(\mathrm{O32}) > 1.0$, more than 60\% are strong LyC leakers ($f_{\rm esc}^{\mathrm{LyC}} > 5\%$). Our stacked spectra show $\log(\mathrm{O32}) > 1.0$ at all radii except for pix$=-2$, where $\log(\mathrm{O32}) = 0.86 \pm 0.05$ (Tab.~\ref{tab:line-ratios}), further supporting the picture of density-bounded or porous ionized regions.
	
	We can also estimate $f_{\rm esc}^{\mathrm{LyC}}$ from the relation between the UV continuum slope and LyC escape fraction derived by \citet{chisholm2022} and calibrated using the LzLCS survey \citep{jaskot2024}. Applying this relation to our radial $\beta_{\rm UV}$ measurements yields $f_{\rm esc}^{\mathrm{LyC}} \sim 3\%$ in the central regions, increasing up to $\sim 15\%$ in the outskirts, in broad agreement with the indirect constraints discussed above. Finally, \citet{schaerer2022} suggested that strong LyC leakers typically exhibit $\CIV/\CIII \gtrsim 0.75$ (see also \citealt{saxena2022}). In our data, \CIV\ is securely detected only in pixels $0$ and $1$, limiting the applicability of this diagnostic across all radii. Nevertheless, the measured ratios are consistent with moderate LyC escape rather than extreme leakage ($f_{\rm esc}^{\mathrm{LyC}} > 0.1$) that would be expected from the observed \CIV/\CIII compared with the threshold defined by \citet{schaerer2022}, as shown in Fig.~\ref{fig:Cdiagn}. Moreover, we find $f_{\rm esc}^{\mathrm{LyC}}<f^{\rm Ly\alpha}_{\rm esc}$, which is consistent with the results reported in \citep{saxena2024}. By means of the same diagnostic plot, our stack is also classified as star-forming lying below the empirical AGN threshold. 
	
	Overall, the combined UV line ratios, O32 values, $\beta_{\rm UV}$ trends, and \lya ~escape fractions—consistently indicate that our LAEs are dominated by stellar photoionization and likely exhibit moderate LyC escape fractions. These properties reinforce the scenario in which typical $z>4$ LAEs represent efficient, though not extreme, contributors to the ionizing photon budget during cosmic reionization.
	
	\section{Constraining feedback models from the SPICE simulations}
	
	In this section, we compare some observed properties of our LAEs stack with predictions from the SPICE simulations. As shown in \citet{Bhagwat2024}, variations in supernova feedback significantly influence the strength of galactic outflows. These differences reshape the interstellar and circumgalactic media, thereby affecting the escape fraction of Lyman-continuum photons and leading to distinct reionization scenarios. The SPICE radiation-hydrodynamic simulations account for two different modes of supernova feedback, namely smooth and bursty \citep{Bhagwat2024, Bhagwat2025}. The difference in the models originates from the timing of injection of supernova energy leading to highly stochastic star-formation histories in the bursty model, while non-stochastic star-formation in the smooth model \citep{Basu2026}. SPICE models the ISM down to $\approx30$ pc (10 pc) at $z=5$ ($z=10$) allowing for self-consistent modeling of nebular lines. We model Ly$\alpha$ and balmer lines assuming collisional and recombination excitations separately assuming a case-B scenario for recombinations (see ~\citealt{Bhagwat2025} for details). Finally, the model to estimate [O III] emission is detailed in \citealt{Casavecchia2026}. For each emission line, we additionally follow the radiative transfer through the multiphase ISM/CGM and dust to account for attenuation and extinction and generate synthetic datacubes of emission lines. To enable a proper comparison with our observed properties, we built a simulated stacked sample, which comprises 132 \lya ~emitters per feedback model in a range of redshift and UV magnitude matching the observed ones. For each LAE, we extracted datacubes in 12 random lines of sight (LOS). Then, for each galaxy and LOS, we placed 5 JWST/NIRSpec mock slits at different inclination angles, all centered on the peak of UV emission. This results in a total of 132$\times$12$\times$5 simulated spectra per feedback model. For each model, we stacked all the generated 2D spectra adopting the same procedure as the observed stack (see Sect.~\ref{sec:method}). Finally, the UV continuum slope, and the EW of \lya, \Halpha and \Hbeta are calculated using the stacked spectra. 
	
	In the top and bottom panels of Fig.~\ref{fig:lya-prop}, we compare the radial profiles of the UV continuum slope, rest-frame EW(\lya), \lya ~escape fraction and metallicity with the results from the SPICE models. The SPICE bursty model produces bluer UV slopes than the smooth model and therefore matches our observations more closely. In both SPICE smooth and bursty models, the EW(\lya) and \lya ~escape fractions increase with radius. However, the bursty model shows a significantly stronger rise, particularly in the second radial bin (pixel 2), where both the EW(\lya) and \lya\ escape fraction increase sharply relative to the central bin. Conversely, the smooth model predicts a more gradual increase with a smaller amplitude for EW(\lya) and a flattening for the \lya\ escape fraction. In SPICE, the enhanced rise shown by the bursty case is associated with feedback driven outflows that reduce central \ion{H}{I} column densities and create low-density channels \citep[see][]{Bhagwat2024}. These ISM conditions increase the escape probability of resonantly scattered \lya\ photons and redistribute them outward relative to the UV continuum. The smooth model instead maintains higher and more centrally concentrated neutral gas columns, leading to lower \lya\ escape fractions and weaker radial variation.
	
	The Balmer line EW profiles provide an independent constraint. Under Case B recombination, \Halpha\ and \Hbeta\ originate from the same H~II regions and trace the same emission measure, with a fixed intrinsic ratio \citep{Osterbrock2006}. Therefore, in the absence of dust, their radial profiles should have identical shapes up to a constant factor. While the observed radial profiles of \Halpha\ and \Hbeta\ equivalent width match and are consistent with negligible dust (see Fig.~\ref{fig:EW-comp}), in the smooth model we find that EW(\Halpha) decreases with radius, but less steeply than EW(\Hbeta). Indeed, the smooth model shows a centrally enhanced dust density, thus suppressing \Hbeta\ more strongly and steepening its radial EW profile relative to \Halpha. 
	
	Finally, the metallicity profiles predicted by SPICE \citep{garcia2025pt1, garcia2025pt2} are also strongly affected by the feedback model (bottom right panel of Fig.~\ref{fig:lya-prop}). The smooth model develops a steep negative metallicity gradient with elevated central metallicities, which is also connected to the stronger central dust attenuation \citep[see][]{Bhagwat2024} and higher attenuation of \Hbeta. Conversely, the bursty model exhibits flatter and overall lower metallicity profiles in agreement with our observed profiles, implying a combination of the redistribution of metal-enriched gas by episodic feedback and inside-out growth. The weaker radial dust gradients in this model also produce more similar EW radial profiles for \Halpha\ and \Hbeta.
	
	Overall, we have demonstrated that comparing synthetic SPICE observables with real data provides a powerful approach to disentangling and constraining different feedback mechanisms at high redshift. Specifically, the combined radial trends in EW(\lya) and \lya ~escape fraction, particularly the pronounced rise at intermediate radii, together with flatter metallicity gradients and moderate Balmer EW slopes, indicate that the bursty SPICE feedback prescription is in agreement with the data presented and thus highly favored.
	
	Such a model naturally provides for a more bursty star formation history, which has been proposed as a potential explanation for the high abundance of luminous sources observed at early cosmic times \citep[e.g.,][]{Gelli2024, Munoz2026}. However, within the SPICE simulations, the dust-attenuated luminosity functions predicted by both the bursty and smooth feedback models reproduce observations equally well up to $z\approx14$ \citep{Bhagwat2024, Basu2026}. This degeneracy suggests that additional constraints are required. In particular, comparisons with other observables, such as galaxy morphology and kinematics, could provide further discriminatory power.

	\section{Summary and Conclusions}
	
	In this work, we performed a spatially resolved stacking analysis of 287 LAEs at $z>4$ using JWST/NIRSpec prism spectroscopy. Specifically, we investigated the radial trend of emission lines and continuum properties on sub-kiloparsec scales. Our main findings can be summarized as follows:
	
	\begin{itemize}
		
		\item LAEs exhibit strong UV and optical emission lines, including Ly$\alpha$, \ion{C}{III}], [\ion{O}{II}], [\ion{Ne}{III}], Balmer lines, and \OIII across multiple spatial elements. High-ionization features such as \CIV and \ion{He}{II} are confined to the central regions, suggesting stratified ionization conditions.
		
		\item Radially increasing blue UV slopes ($\beta_{\rm UV}\sim -2.2$ on average) and Balmer decrements consistent with Case~B recombination indicate negligible dust attenuation at all radii.
		
		\item Using $T_e$-based measurements, \textsc{HCm}, and photoionization models, we consistently derive sub-solar metallicities of $\sim10$--15\% $Z_\odot$ ($12+\log({\rm O/H})\sim7.5$--7.9). We find evidence for either a negative (i.e. decreasing) or flat metallicity gradient with radius. The ionization parameter is high, $\log(U)\sim -1.5$ in the central regions, mildly decreasing outward, and the gas density is consistent with $\log(n/{\rm cm}^{-3})\sim3$.
		
		\item We measure elevated nitrogen-to-oxygen ratios ($\log({\rm N/O}) \sim -0.4\pm 0.3$), placing LAEs among the growing population of nitrogen-enriched galaxies at high redshift.
		
		\item While non-resonant optical lines show declining equivalent-width profiles with radius, EW(Ly$\alpha$) increases toward the outskirts, indicating spatially extended Ly$\alpha$ emission relative to the UV continuum. The Ly$\alpha$ escape fraction rises from $\sim16\%$ in the center to $\gtrsim24\%$ in the outer regions.
		
		\item We infer $\log(\xi_{\rm ion}/{\rm Hz\,erg^{-1}})\simeq25.2\pm0.1$, consistent with galaxies of similar UV luminosity and supportive of efficient ionizing photon production.
		
		\item Comparison with SPICE radiation-hydrodynamic simulations favors bursty supernova feedback models, which better reproduce the observed radial trends of EW(Ly$\alpha$) and $f_{\rm esc}^{\rm Ly\alpha}$.
		
	\end{itemize}
	
	Overall, our analysis demonstrates that Ly$\alpha$ emission is not merely a tracer of instantaneous star formation, but a sensitive probe of the internal structure, feedback processes, and circumgalactic environment of early galaxies. The increasing Ly$\alpha$ escape fraction with radius supports a scenario in which resonant scattering redistributes Ly$\alpha$ photons into outer regions, where they can escape more efficiently. These properties suggest that typical $z>4$ LAEs are efficient, though not extreme, contributors to the ionizing photon budget during cosmic reionization.
	
	Importantly, our stacking analysis probes the bulk of the LAE population at $z_{\rm median}=5.5$, characterized by $M_{\rm UV, median}=-18.7$ and EW(\lya)$_{\rm median}$=61~\AA, rather than rare, exceptionally luminous systems. Our results therefore provide a representative view of the physical conditions and Ly$\alpha$ radiative transfer properties in typical high-redshift LAEs.
	
	More broadly, this work highlights the power of stacked JWST spectroscopy to deliver a population-level view of the internal structure of early galaxies that would otherwise remain inaccessible. By statistically resolving radial trends in emission-line and continuum properties, stacking bridges the gap between detailed studies of rare bright systems and the average properties of the high-redshift galaxy population. As JWST surveys continue to expand and the number of spectroscopically confirmed LAEs increases, this approach will enable us to explore how Ly$\alpha$ emission, metallicity, ionization conditions, and escape fractions depend on $M_{\rm UV}$, redshift, and EW(\lya), providing a more comprehensive framework for understanding the role of LAEs during the first billion years of cosmic history.
	
	\begin{acknowledgements}
		We thank the anonymous referee for the insightful comments that improved our work. We acknowledge support from PRIN 2022 MUR project 2022CB3PJ3 - First Light And Galaxy aSsembly (FLAGS) funded by the European Union – Next Generation EU and from the ERC synergy grant 101166930 - RECAP.\\
		This work is based on observations made with the NASA/ESA/CSA James Webb Space Telescope, obtained at the Space Telescope Science Institute, which is operated by the Association of Universities for Research in Astronomy, Incorporated, under NASA contract NAS5-03127.
		Support for program number GO-6368 was provided through a grant from the STScI under NASA contract NAS5-03127. The data were obtained from the Mikulski Archive for Space Telescopes (MAST) at the Space Telescope Science Institute. These observations can be accessed via \href{http://dx.doi.org/10.17909/0q3p-sp24}{doi:10.17909/0q3p-sp24}.
		
		Some of the data products presented in this work were retrieved from the Dawn JWST Archive (DJA). DJA is an initiative of the Cosmic Dawn Center (DAWN), which is funded by the Danish National Research Foundation under grant DNRF140.
	\end{acknowledgements}
	
	\bibliography{biblio}{}
	\bibliographystyle{aa}
	
	\begin{appendix}
		
		\nolinenumbers
		
		\section{Comparison between the CAPERS and DJA spectra}
		\label{app:capers}
		
		\begin{figure}
			\centering
			\includegraphics[width=0.9\linewidth]{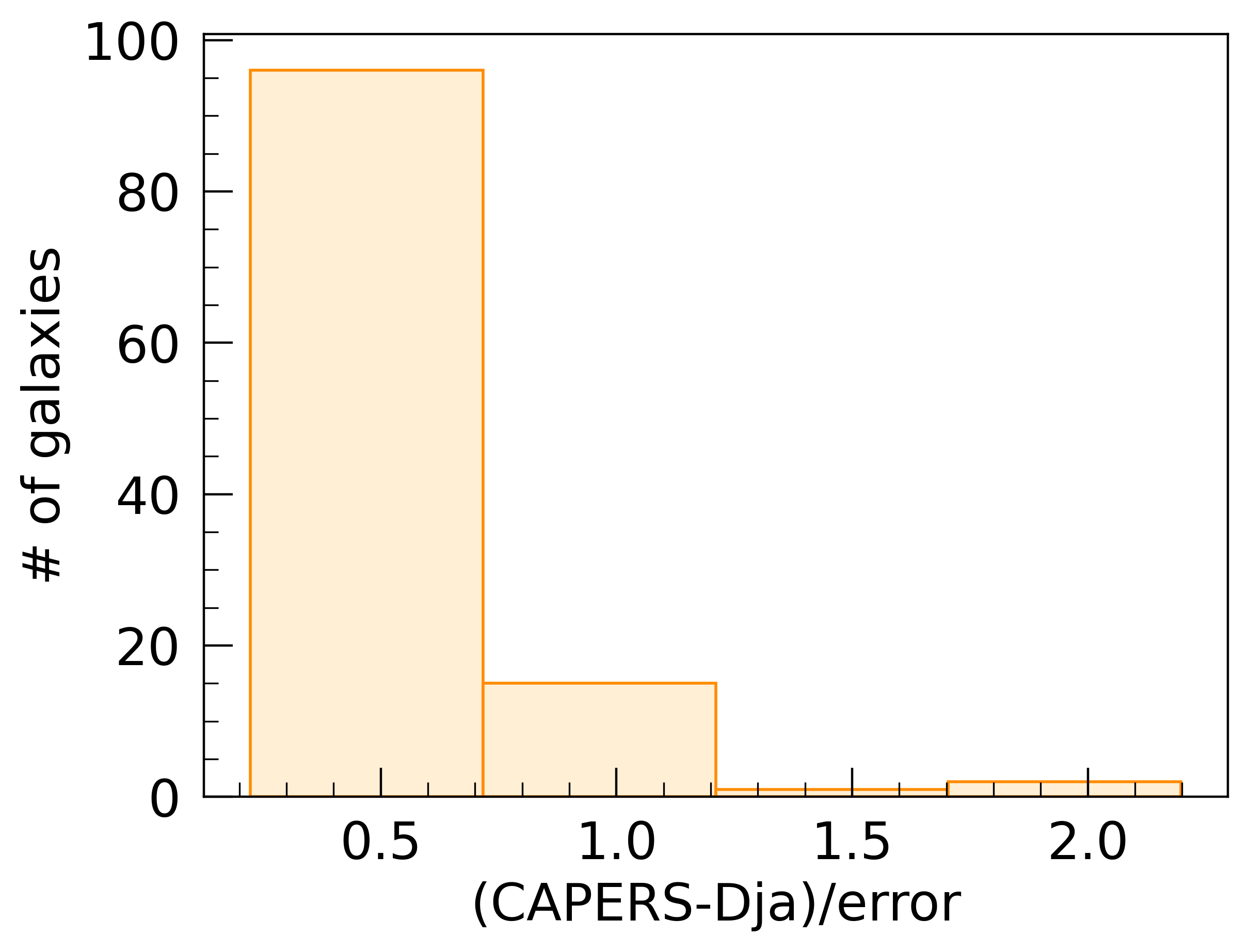}
			\caption{Distribution of the median difference between CAPERS and DJA spectra, as defined in Appendix \ref{app:capers}.}
			\label{fig:app-comparison-spec}
		\end{figure}
		
		To ensure a uniform and self-consistent framework, we checked the consistency between CAPERS and DJA pipelines for the $\sim 100$ galaxies analyzed using the CAPERS-reduced spectra. For each of these galaxies, we compared the CAPERS 2D-spectrum pixel by pixel with its DJA 2D-spectrum. The wavelength grid is almost the same in the two reductions: the DJA one is just slightly more extended in the red part of a few hundred angstrom, but the binning is almost consistently the same. For each galaxy, we computed the median value of the difference between the CAPERS and DJA spectra at each wavelength as ${\rm diff} = |{\rm flux_{\rm \lambda,CAPERS}} - {\rm flux_{\rm \lambda,DJA}}|/{\rm error}$, where error = $\sqrt{\Delta{\rm flux_{λ,CAPERS}}^2 + \Delta{\rm flux_{λ,DJA}}^2}$. In Fig.\ref{fig:app-comparison-spec}, we report the final distribution of the median diff for the pixel 0. As seen from the figure, 97\% of the spectra match within 1$\sigma$ error. The percentage increases to 100\% going towards external pixels (either up or downwards). Therefore, any differences in the data reduction do not impact our results on the stacking analysis.
		
		\section{Effects of the wavelength-dependent PSF in stacking}
		\label{sec:app-check-PSF}
		
		\begin{figure}
			\centering
			\includegraphics[width=0.95\linewidth]{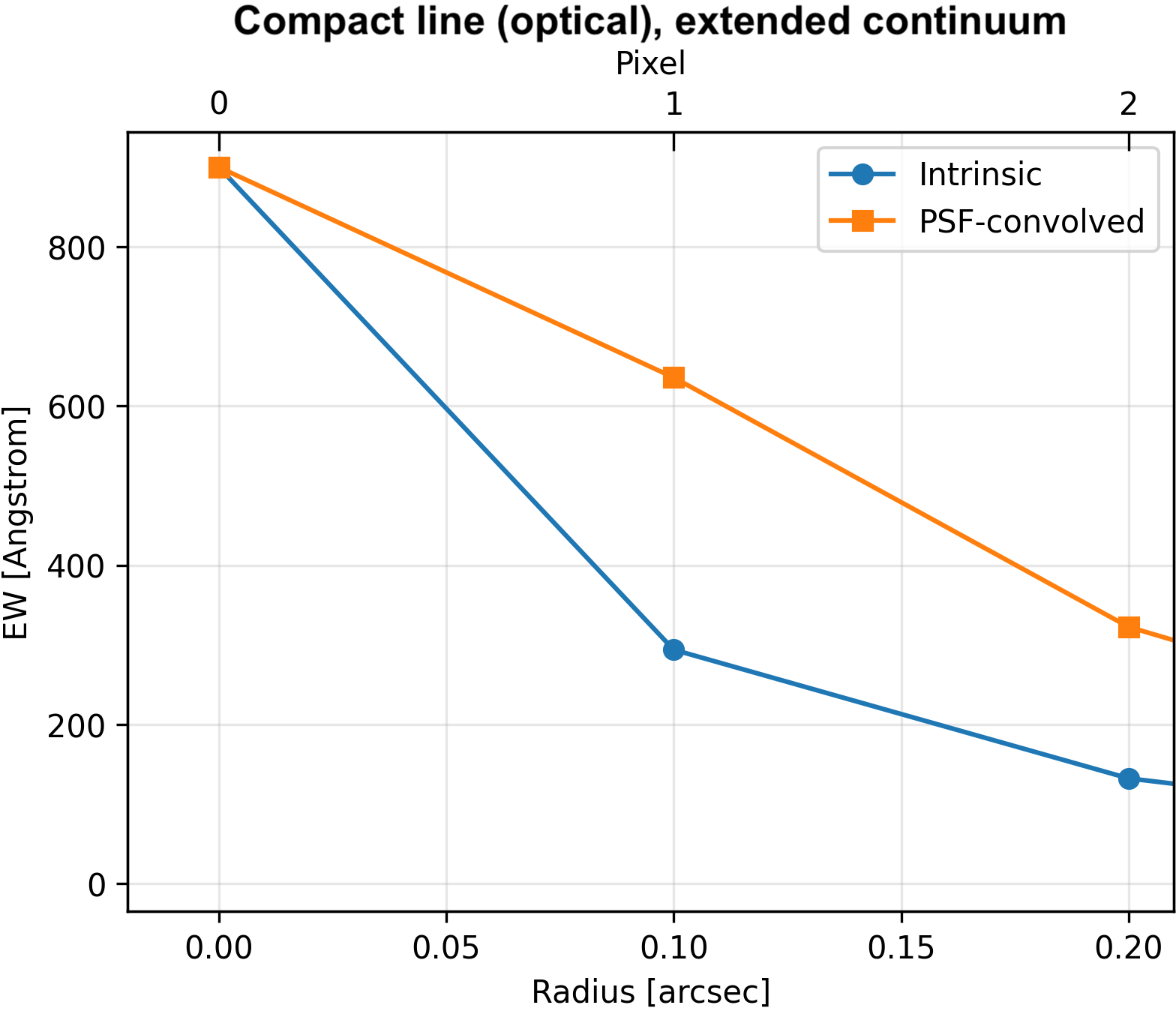}\\[0.2cm]
			\includegraphics[width=0.95\linewidth]{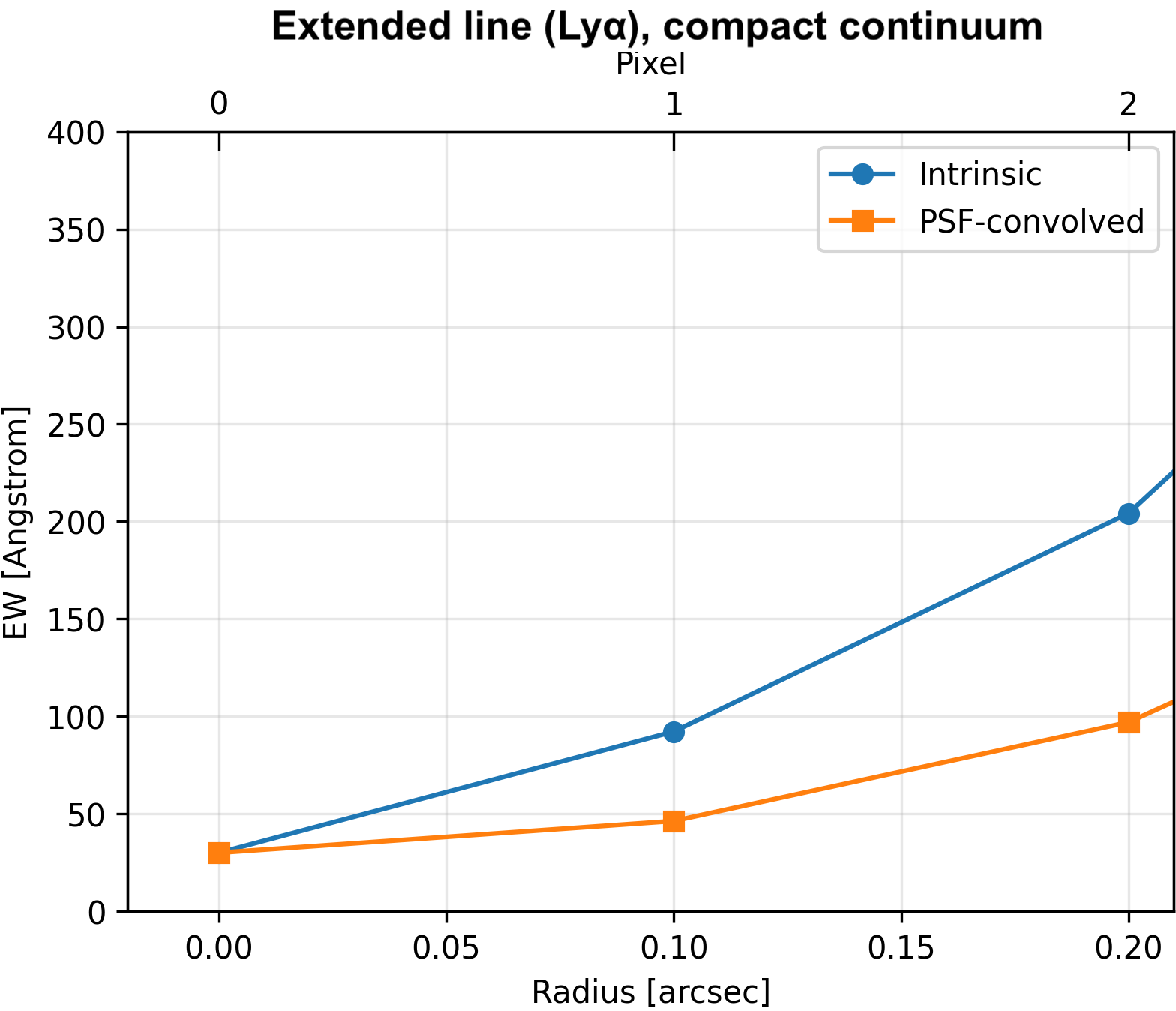}\\[0.2cm]
			\includegraphics[width=0.95\linewidth]{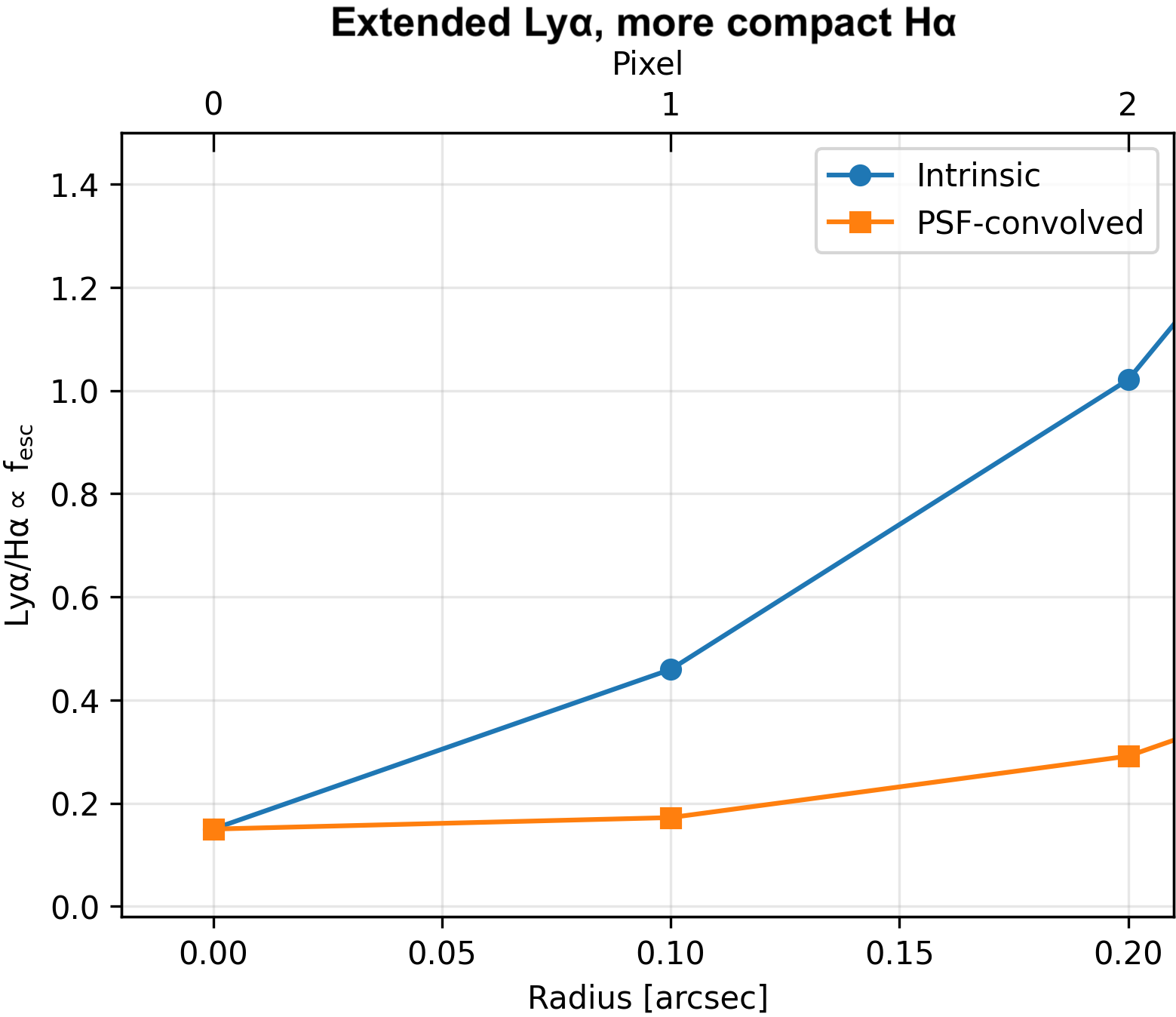}\\
			\caption{Tests on the radial profiles extracted from 2D spectroscopy. Intrinsic (blue) and PSF-convolved (orange) radial profiles assuming the line emission to be more compact than the continuum at \OIII wavelength (top panel); the line emission more extended than the continuum at \lya~ wavelength (central panel); the ratio between extended \lya~ and compact \Halpha line emission profiles (bottom panel). Details on the specific assumption for the line and continuum profiles are reported in the main text.}
			\label{fig:app-EW-profile-sim}
		\end{figure}
		
		 We report the tests performed on simulated data to check the effect of the wavelength dependence of the NIRSpec/prism PSF on the radial trends inferred from 2D spectroscopy, assuming different combinations of morphologies for line and continuum emission. The variation of the effective PSF size with wavelength has been obtained in a previous study involving NIRSpec MSA/prism spectroscopy (see Fig. A.2 in \citealt{Tripodi2024c} for NIRSpec/prism in MSA; also \citealt{deugenio2024} for NIRSpec/prism in IFU). The PSF size changes as follows: $\sigma =0.045"$ (i.e. $\sim0.9$ pixel in the 2D MSA spectrum) at 0.8$\mu$m, $\sigma =0.055"$ at 3$\mu$m, $\sigma=0.065"$ (i.e. 1.3 pixel in the 2D spectrum) at 4$\mu$m, which correspond to the wavelengths of \lya, \OIII and \Halpha at $z=5.5$ (i.e., the median redshift of our sample). 
			
			From recent morphological studies of high-z galaxies in COSMOS, EGS, UDS \citep[e.g.][]{allen2025, ward2024}, the effective radius of the galaxy continuum emission is $\sim 0.1-0.4$ arcsec, with little variation between UV and optical at $z>4$, and a Sérsic index $n\sim1.3$, consistent with disk-like morphologies. Thus, based on these findings, we performed the following tests, assuming that the line emission can be more extended or less extended than the continuum emission. Considering the median $z$ of our sample, in top panel of Fig.~\ref{fig:app-EW-profile-sim}, we compare the intrinsic EW profiles with the PSF-convolved EW profiles for a line emission peaking at observed $\sim$3$\mu$m (i.e. \OIII), assuming the continuum profile to be a Sérsic profile with $n=1.3$ and $R_e=0.25$ arcsec (changing this up to 0.4 arcsec does not affect the outcome of the following tests), and the line emission to be more compact, i.e. for instance a Sérsic profile with $n=1.3$ and $R_e=0.10$ arcsec. The EW profiles have been normalized at $r=0$ arcsec to 900 \AA to match our EW(\OIII) in pixel 0. The PSF convolution does not change the overall decreasing intrinsic trend (blue line), while just making the profile shallower (orange line). The same applies if considering a line and continuum at $\sim$4$\mu$m (corresponding to \Halpha at $z\sim 5.5$).
			
			Analogously, we considered the case of having a line emission more extended than the continuum, as for the case of \lya. In the central panel of Fig.~\ref{fig:app-EW-profile-sim}, we considered a line profile having $n=1.3$ and $R_e=0.25$ arcsec, and the continuum as $n=1.3$ and $R_e=0.10$ arcsec. We normalized the profiles at $r = 0$ at the value of the observed EW(\lya)=30 \AA. Also in this case, the observed profile is shallower than the intrinsic one, while the overall increasing trend is preserved.
			
			Finally, we checked the effect of comparing lines emitting at different wavelengths, as when dividing the \lya~ flux over \Halpha to derive the escape fraction of \lya. In this case, indeed, the two line profiles are convolved with different PSF profiles ($\sigma =0.045"$ at 0.8$\mu$m, $\sigma=0.065"$ at 4$\mu$m). Assuming that the \lya~ emission is more extended than the \Halpha one, i.e. having $n=1.3$ and $R_e=0.25$ arcsec for the former, and $n=1.3$ and $R_e=0.10$ arcsec for the latter, we obtain the profiles shown in the bottom panel of Fig.~\ref{fig:app-EW-profile-sim}.
			
			We normalize the profiles at $r = 0$ to the observed value of escape fraction, i.e. $f_{\rm esc}=0.15$. Interestingly, also in this case, the difference of a factor 1.4 in the PSF size at \lya~ and \Halpha wavelengths is not sufficient to change the overall radial trend, while still making the profile shallower. Interestingly, our estimate of the escape fraction as presented in this work can be considered as a lower limit at all radii, especially in the outer regions. 
			
			Overall, we conclude that the observed radial trends (either in EW or $f_{esc}$) are driven by the physical radial profiles of the line or continuum emission rather than PSF convolution effects or differences in the PSF sizes. A precise quantification of the difference between the intrinsic and convolved profiles depends on the assumptions for the specific shape of the line and continuum emission profiles. Indeed, there are combinations of $R_e$ (and also $n$) for which the effect of the PSF convolution is minimized if not negligible. However, restricting to the assumptions used in these tests, we may say that for the first case, the PSF convolution alters the observed value of the EW by a factor of 2, at most, in the central region that decreases towards the outskirts, while for the other two tests the effect is smaller in the center (a factor of 2) and increases towards the outskirts.

		\section{Table}
		\label{app:tables}
		
		In Tab.~\ref{tab:line-ratios} we report the line ratios used in Sects.~\ref{sec:contam}, \ref{sec:N2Ha-ratio} to estimate the contamination correction for \NeIIIe$_{\rm blend}$ and the contribution of \NII to \Halpha, respectively.
		
		\begin{table*}[]
			\centering
			\caption{Line ratios}
			\begin{tabular}{c|ccccc}
				Line ratio & Pix -2 & Pix -1 & Pix 0 & Pix 1 & Pix 2 \\
				\hline
				\Hdelta/\NeIIIe$_{\rm blend}$ & $0.42\pm 0.20$ & $0.34\pm 0.04$ & $0.37\pm 0.03$ & $0.30\pm 0.04$ & $0.34\pm 0.10$ \\
				\Heta/\NeIIIe$_{\rm blend}$ & $0.11\pm0.04$ & $0.10\pm 0.01$ & $0.10\pm0.01$ & $0.08\pm0.01$ & $0.10\pm 0.02$ \\
				\Hzeta/\NeIIIe$_{\rm blend}$ & $0.17\pm0.05$ & $0.14\pm0.02$ & $0.15\pm0.01$ & $0.12\pm 0.02$ & $0.14\pm 0.04$ \\
				\HeIL/\NeIIIe$_{\rm blend}$ & N/A & $0.24\pm 0.03$ & $0.20\pm 0.02$ & $0.17\pm 0.03$ & N/A\\
				\NeIII/\NeIIIe$_{\rm blend}$ & $<0.7$ & $0.52\pm 0.06 $ & $0.55\pm 0.04$ & $0.63\pm 0.06$ & $<0.8$ \\
				\NII/\Halpha & N/A & $0.10\pm 0.05$ & $0.10\pm 0.02$ & $0.17\pm 0.05$ & N/A\\
				\OIIIall/\OIIall & $12_{-2}^{+3}$ & $11\pm 1$ & $15\pm1$ & $12\pm 1$& $7\pm 1$ \\[0.1cm]
				\hline
				\hline
			\end{tabular}
			\tablefoot{Columns: ratio names; values of line ratio in different pixels from -2 to 2, assuming no dust correction.}
			\label{tab:line-ratios}
		\end{table*}
		
		\section{Comparison of the models with and without the [\ion{N}{II}] component}
		\label{app:NIIcomp}
		
\begin{figure}
			\centering
			\includegraphics[width=0.9\linewidth]{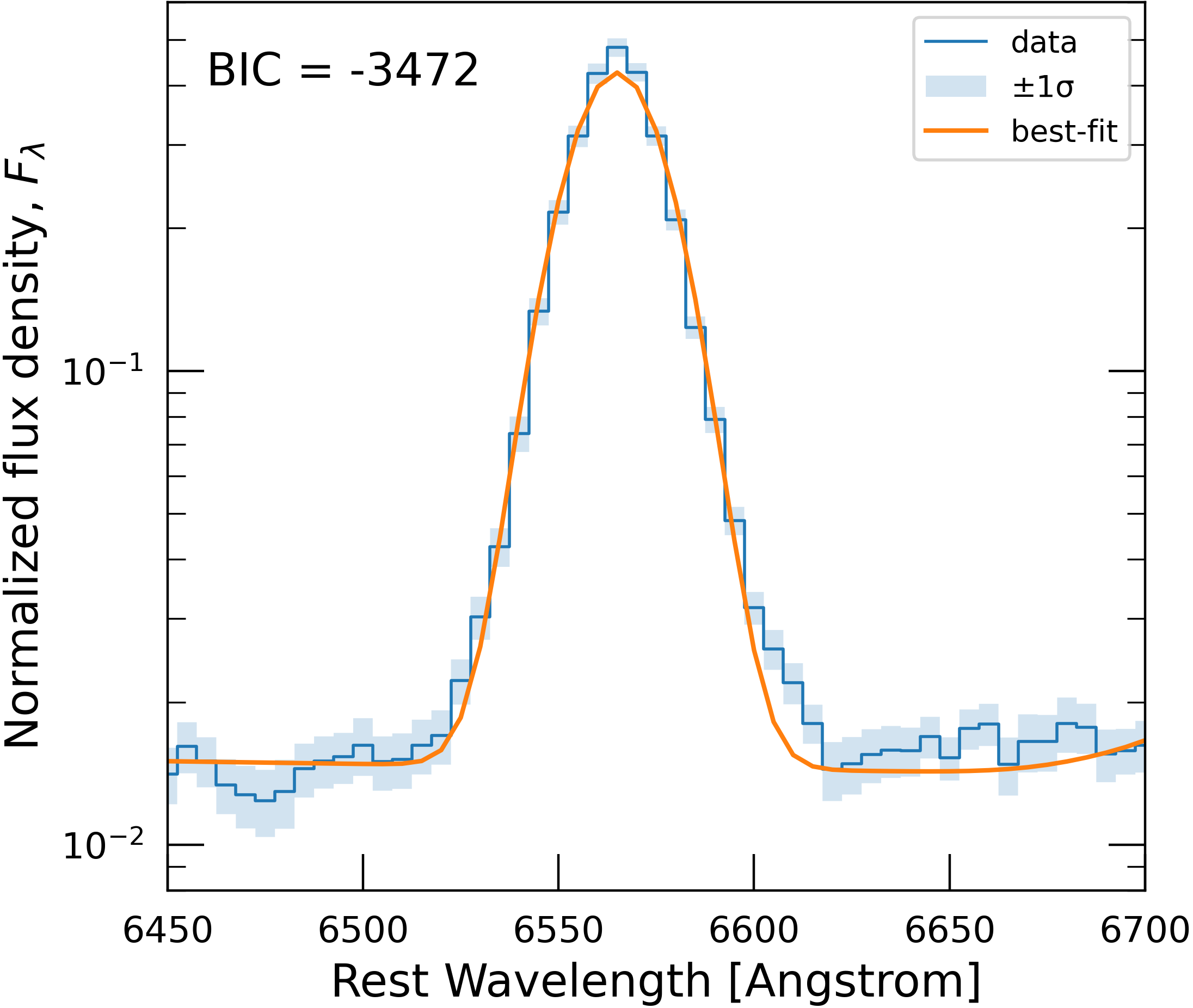}\\[0.1cm]
			\includegraphics[width=0.9\linewidth]{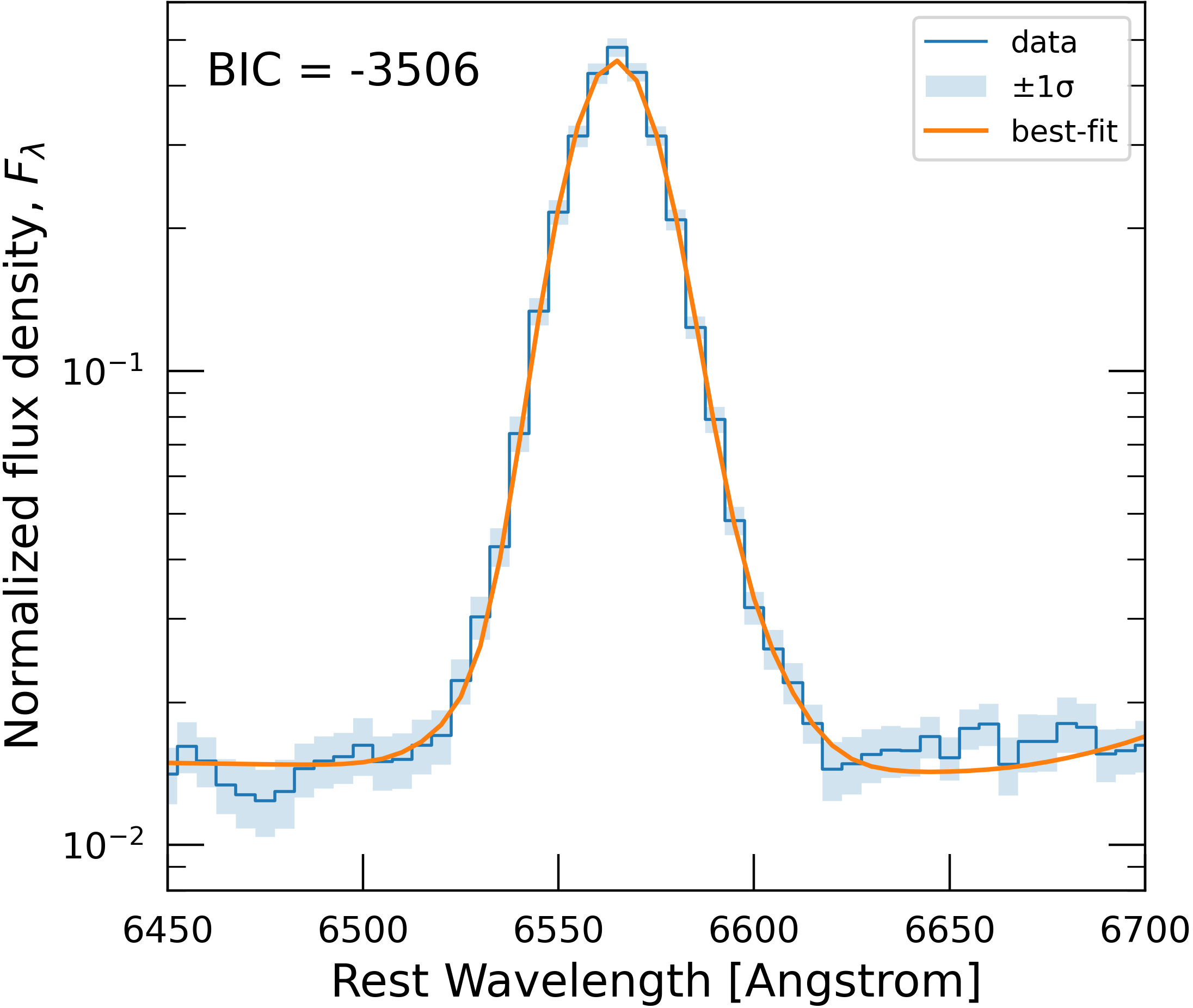}
			\caption{Comparison of the best-fitting models of the blended \Halpha-\NIIall without (top) and with (bottom) the [\ion{N}{II}] component.}
			\label{fig:HaNII-fit}
		\end{figure}

		In Fig.~\ref{fig:HaNII-fit}, we compare the best-fitting models of the blended \Halpha–\NIIall feature obtained without (top panel) and with (bottom panel) the inclusion of the [\ion{N}{II}] component. The observed spectrum clearly favors the model including the [\ion{N}{II}] doublet, as it accounts for the excess flux on the blue side of \Halpha and yields a significant improvement in the Bayesian Information Criterion (BIC).
		
		\section{Spectra for off-centered pixels}
		\label{app:spectra}

		\begin{figure*}[t]
			\centering
			\includegraphics[width=0.9\linewidth]{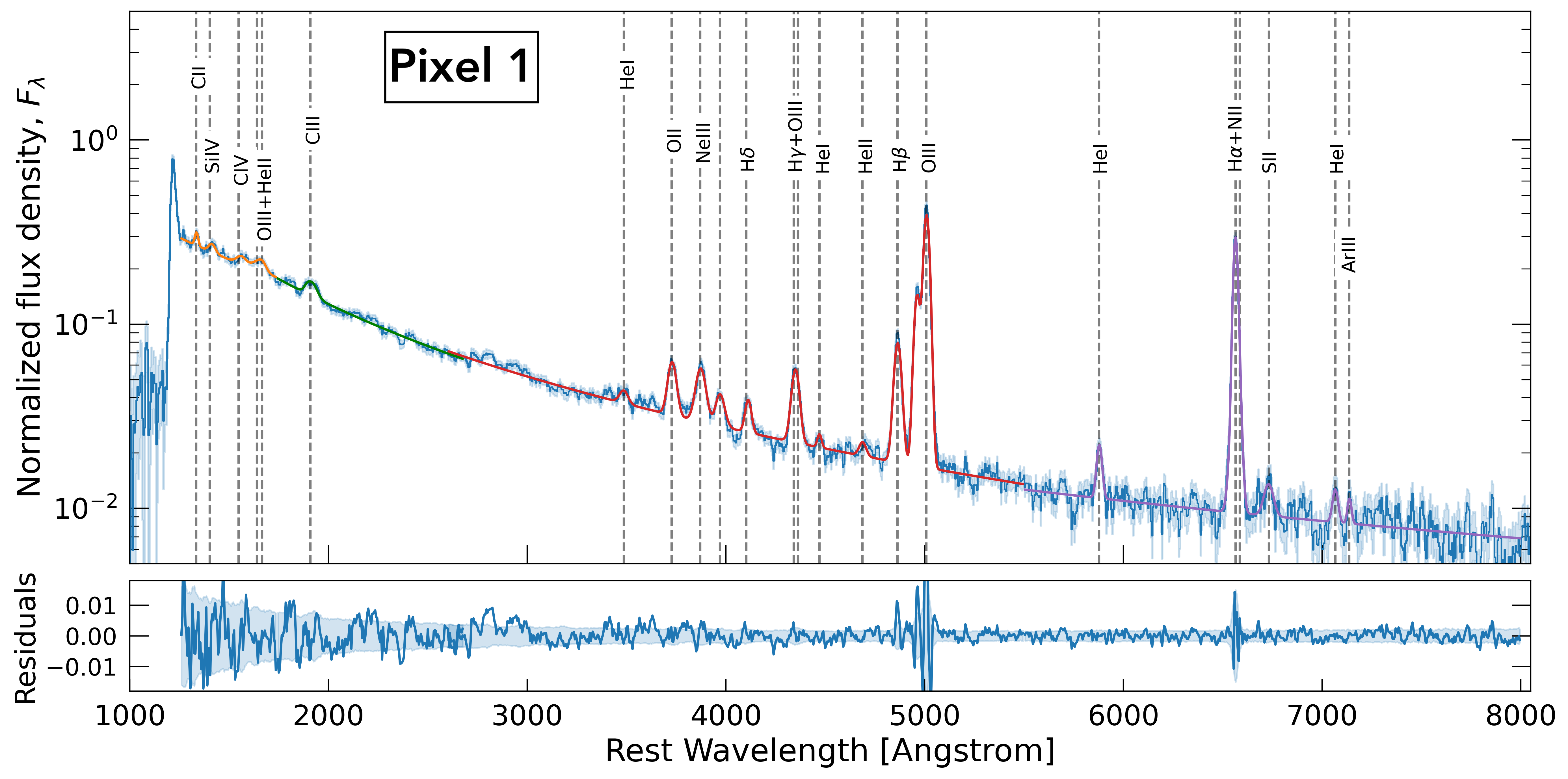}
			\includegraphics[width=0.9\linewidth]{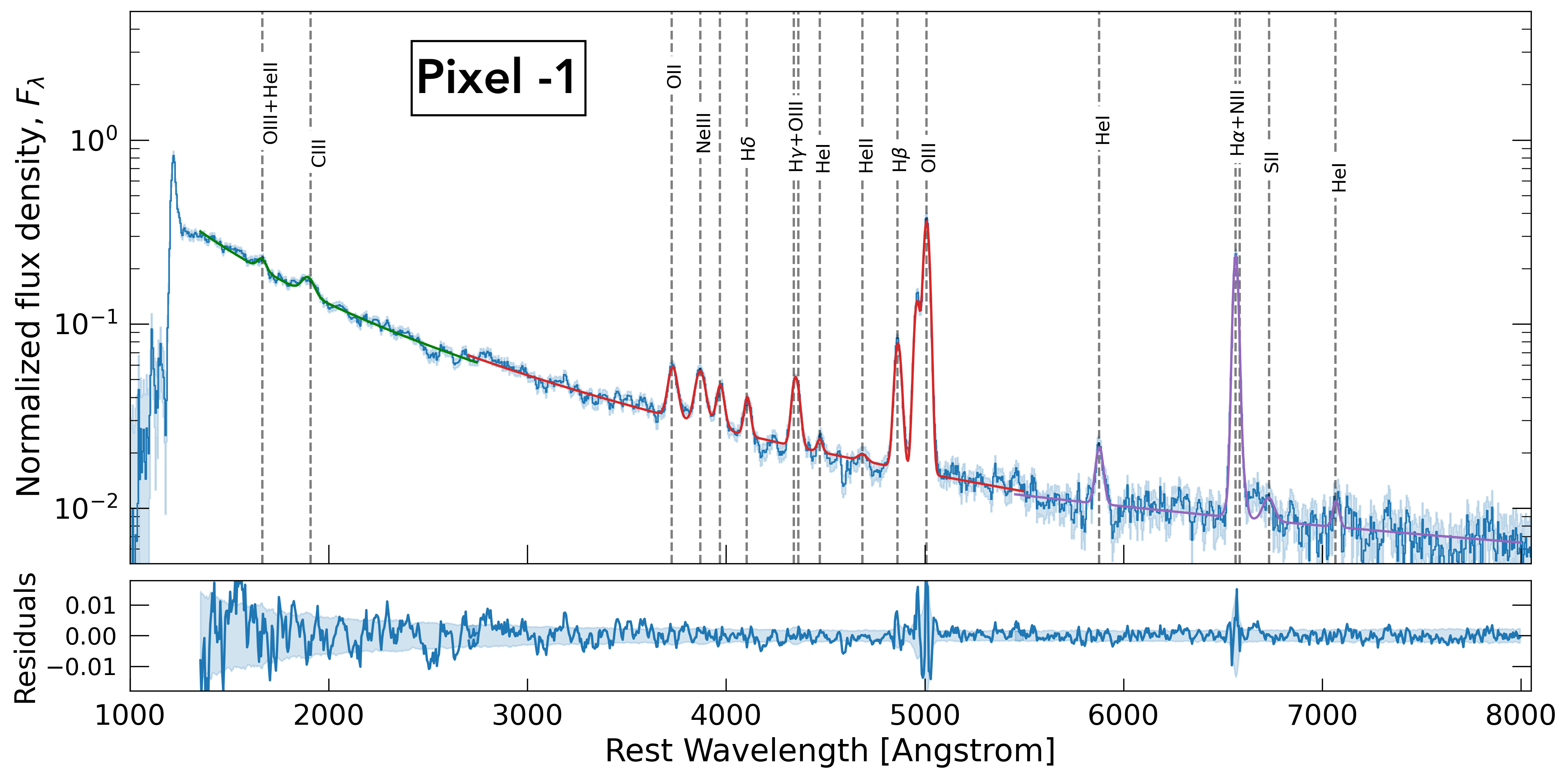}
			\caption{Same as Fig. \ref{fig:fit-spec-pix0} for pixels 1 and -1 of the LAEs stack.}
			\label{fig:fit-spec-pix1}
		\end{figure*}
		
		\begin{figure*}
			\centering
			\includegraphics[width=0.9\linewidth]{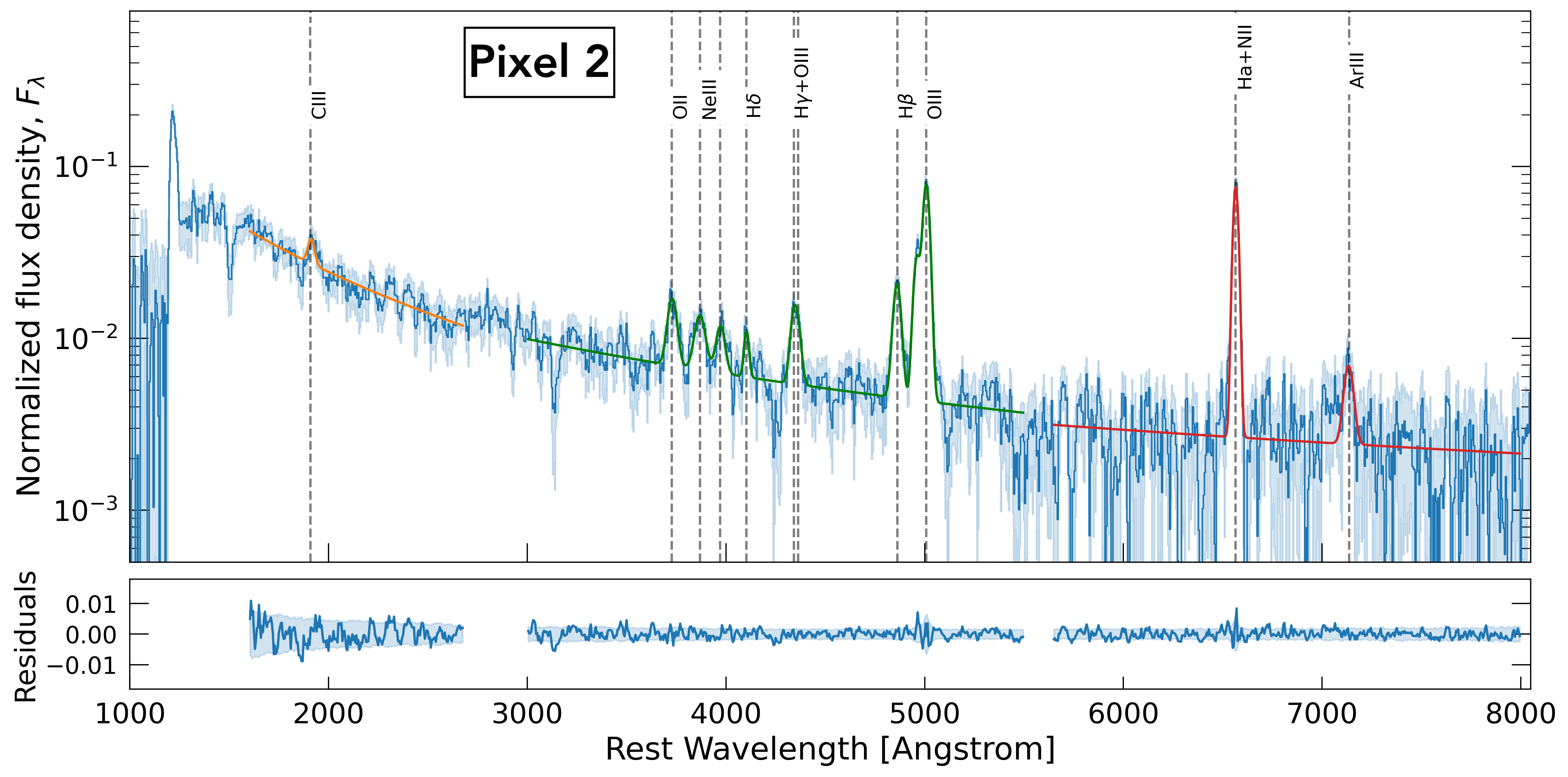}
			\includegraphics[width=0.9\linewidth]{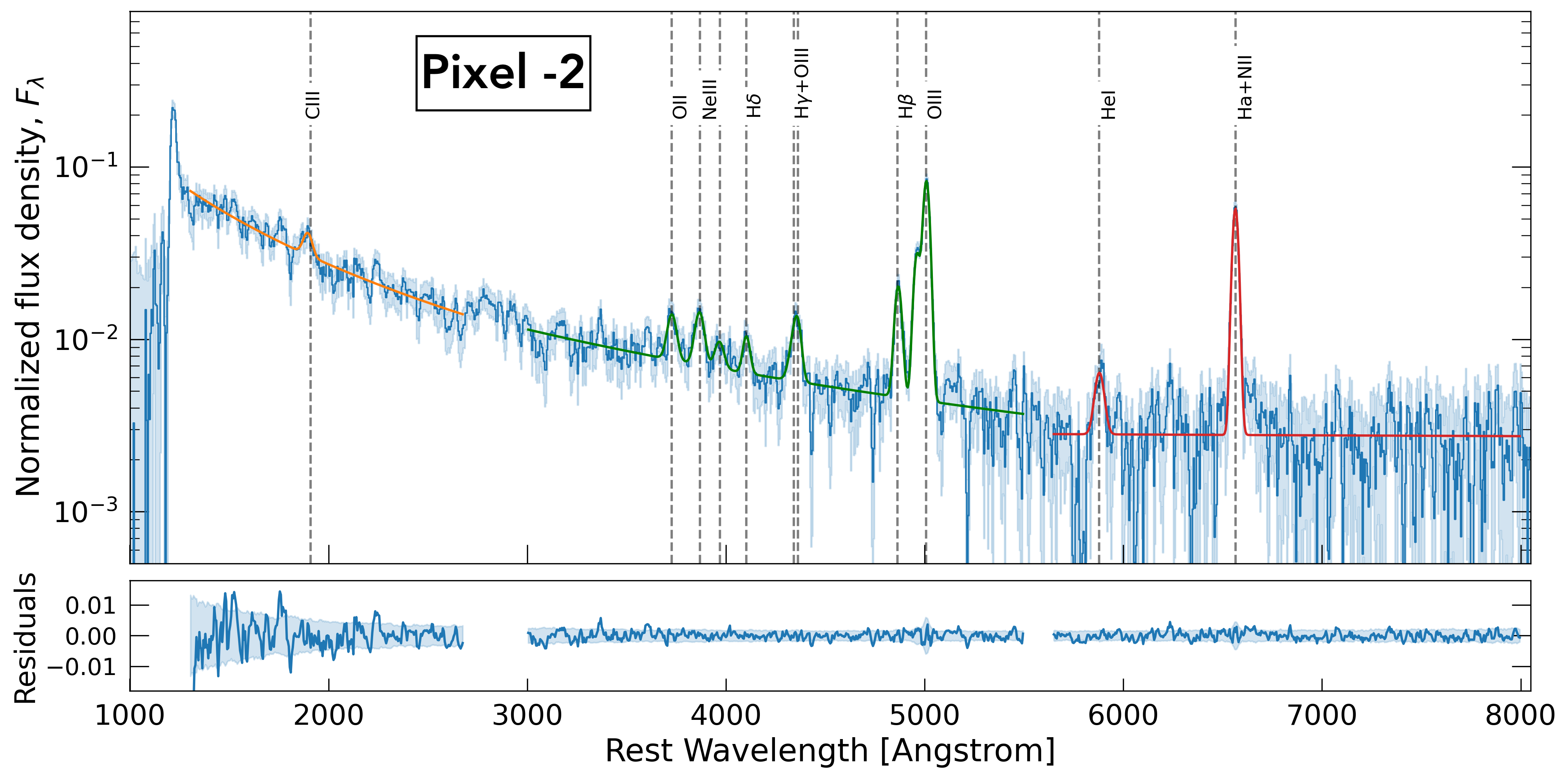}
			\caption{Same as Fig. \ref{fig:fit-spec-pix0} for pixels 2 and -2 of the LAEs stack.}
			\label{fig:fit-spec-pix2}
		\end{figure*}
		
		In Fig.~\ref{fig:fit-spec-pix1}-\ref{fig:fit-spec-pix2}, we show the spectra extracted from the off-centered pixels (pixels 1,-1,2,-2).

		\section{OHNO diagram with conservative blending correction for \NeIIIe}
		\label{app:fig-OHNO}
		
\begin{figure}
			\centering
			\includegraphics[width=0.9\linewidth]{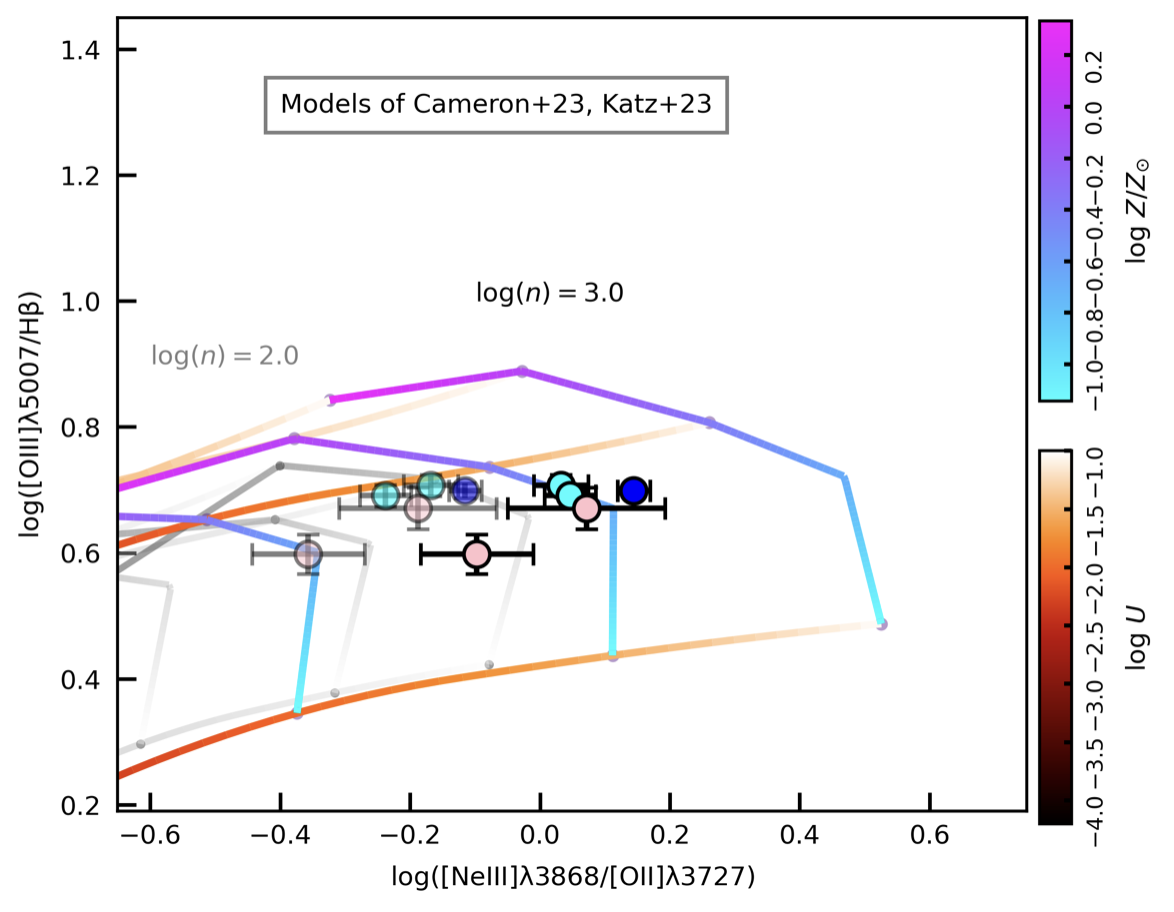}
			\caption{Same as the right panel of Fig.~\ref{fig:OHNO} but with the \NeIIIe/\OII ratio corrected for the blending contamination as lighter circles. The maximum correction for \NeIII due to the blend with \HeIL, \Heta and \Hzeta is $\approx$ 0.3 dex. }
			\label{fig:app-corr-Ne3}
		\end{figure}

		In Fig.~\ref{fig:app-corr-Ne3} we show the \NeIIIe/\OII ratio corrected for the blending contamination along with the uncorrected ratios and the photoionization models, already presented in Fig.~\ref{fig:OHNO}.
		
	\end{appendix}
	
\end{document}